\documentclass[aps,prb,twocolumn,superscriptaddress,floatfix,citeautoscript,hyperref=pdftex]{revtex4-1}

\usepackage{verbatim}
\usepackage{latexsym,amssymb,float}
\usepackage{setspace}
\usepackage{graphicx}
\usepackage{dcolumn}
\usepackage{epsfig}
\usepackage{amsmath}
\usepackage[colorlinks=true,linkcolor=red,urlcolor=purple,citecolor=blue]{hyperref}
\usepackage{bm}
\usepackage{tikz}
\usepackage{caption}
\usepackage{subcaption}
\normalsize 
\usepackage{booktabs}
\usepackage{braket}
\usepackage{times}
\usepackage{bbold}

\newcommand{\be}{\begin{equation}}
\newcommand{\ee}{\end{equation}}
\renewcommand{\vec}[1]{\bm{#1}}

\begin{document}

\title{Charge-density wave fluctuation driven composite order in the layered Kagome Metals}
\author{Alexei M. Tsvelik}
\author{Saheli Sarkar}

\affiliation{
    Division of Condensed Matter Physics and Materials Science, Brookhaven National Laboratory, Upton, NY 11973-5000, USA}

\begin{abstract}
The newly discovered kagome metals AV$_3$Sb$_5$ (A = K, Rb, Cs) offer an exciting route to study exotic phases arising due to interplay between electronic correlations and topology. Besides superconductivity, these materials exhibit a charge-density wave (CDW) phase occurring at around 100 K, whose origin still remains elusive. The robust multi-component $2 \times 2$ CDW phase in these systems is of great interest due to the presence of an unusually large anomalous Hall effect. In quasi-2D systems with weak inter-layer coupling fluctuation driven exotic phases may appear. In particular in systems with multi-component  order parameters fluctuations may lead to establishment of composite order when only products of individual order parameters condense while the individual ones themselves remain disordered. We argue that such  fluctuation-driven regime of composite CDW order  may exist  in thin films of kagome metals above the CDW transition temperature. It is suggested that the melting of the Trihexagonal state in the material doped way  from the van Hove singularities gives rise to a pseudogap regime  where the spectral weight is concentrated in small pockets and  most of the original Fermi surface is gapped. Our findings suggest possible presence of exotic phases in the weakly coupled layered kagome metals, more so in the newly synthesized thin films of kagome metals.  

\end{abstract}


\maketitle

\section{Introduction}
The interplay between electronic correlations and topology is a major field of study in the condensed matter systems \cite{kennes2021moire,dzero2016topological}. The recently discovered kagome metals AV$_3$Sb$_5$ with (A = K, Rb, Cs) are quasi-two dimensional (2D) system with hexagonal lattice symmetry \cite{BOrtizPRM2019}. The band structure of the kagome metals exhibit a flat band, saddle-point van Hove singularities (vHSs) and a pair of Dirac points. Owing to such an electronic structure, these systems have created a new platform to study exotic phases which can occur due to presence of both correlations and topology \cite{ortiz2020cs,Ortiztopometal2021}.

All of the AV$_3$Sb$_5$ undergo a charge-density-wave (CDW) transition \cite{jiang2021unconventional,LiPRXphonon2021,uykur2022optical,OrtizCDWPRX2021,TanPRL2021} at around temperature T$_{CDW} \sim$ 100 K. Along with the emergence of the CDW order, experiments and theoretical studies have found different unusual properties, such as bond density modulations \cite{MDennerbondPRL2021}, a chiral flux phase \cite{feng2021chiral,yu2021evidence}, a giant anomalous Hall effect \cite{yang2020giant,FHYuPrb2021,kenney2021absence} with  time-reversal symmetry breaking \cite{mielke2022time,KhasanovPRRTR2022, gupta2022two,xu2022universal}, which can be associated with  loop currents \cite{LinLoopprb2021,ChristensenloopPRB2022,wang2020proximity}.

At much lower temperatures these materials may exhibit superconductivity \cite{ortiz2020cs,KChenPRL2021,ni2021anisotropic,mu2021s} with T$_c \sim $ 1 K. The nature of the superconducting phase is still under debate. Some experiments found the gap to be  nodeless \cite{duan2021nodeless}, some to contain  nodes \cite{zhao2021nodal}. Theoretical studies suggest unconventional nature of the superconductivity\cite{MKieselPRL2013,WangPRB2013,WuPRL2021,WenPRB2022,lin2022multidome}. There have been also proposals of more exotic superconductivity like pair-density wave \cite{chen2021roton,zhou2022chern}, charge 4e and charge 6e superconducting states \cite{zhou2022chern} and nematicity \cite{xiang2021twofold,nie2022charge,grandi2023theory}. 

There have been a great amount of works \cite{kang2022twofold,Louupeakdiphumpprl2022,luo2022electronic,Wuprb2022,tazai2022mechanism,feng2023commensurate} to gain insight into  the nature of the CDW phase. So far it is well established that the CDW order of the kagome metals is a multicomponent (3Q) one, although, the real space structure of the CDW phase still remains elusive. Experiments \cite{luo2022possible,HuPRB2022} observe both Star of David (SoD) and Trihexagonal (TrH) pattern in the two-dimensional plane of these systems. Moreover, the CDW order doubles the unit cell in the (a,b) plane and hence has a robust $2 \times 2$ feature as found in scanning tunneling-microscopy (STM) \cite{jiang2021unconventional}, angle-resolved photoemission spectroscopy (ARPES) \cite{ChoARPESPRL2021} and X-ray \cite{li2022conjoined} experiments. However, some X-ray and STM experiments found a modulation in the crystallographic c- direction for the  kagome metals with alkali atoms Rb and Cs. The simultaneous ordering of CDW phase with commensurate momenta 3Q are believed to be driven by nested Fermi surface instabilities \cite{BalentsPRB2021,nandkishore2012chiral,MKieselPRL2013,WanginstPRB2013}, enhanced through the presence of vHS due to logarithmically diverging density of states at the vHS points \cite{VanHove1953} in two dimensions.  In this paper we explore the situation \cite{MKieselPRL2013} of 5/12 filled band when the chemical potential lies at the van Hove singularity. 

According to the Mermin -Wagner theorem \cite{MerminWagnerprl1966} fluctuations are enhanced in low dimensions. The presence of strong fluctuations  is well established in such   quasi-2D systems  as cuprates, iron based superconductors \cite{Fernandesnematicprb2012} where they are responsible for  pseudogap phase \cite{VarmaPRB2006,PALeePRX2014,pepin2020fluctuations}, anomalous phonon softening \cite{Sarkarphonon2021} and also different emergent orders \cite{Wangnematicprb2014,tsvelik2014composite,sarkarloopcurrentprb2019}. These prototype examples indicate that fluctuations may also play an important role  in the layered quasi-2D kagome metal materials. However, as of now, although several theoretical works have considered a mean-field scenario of the CDW order parameters, the effect of fluctuations in kagome CDW metals has not been discussed. Their  effect will become even more important in the kagome metal mono-layers \cite{kim2023monolayer} and  thin films \cite{Songprl2021,song2021competing,wang2021enhancement}. 

In this paper, we go beyond the mean-field theory of the multi-component CDW order and consider the fluctuations in these orders within a Ginzburg-Landau (GL) free energy model. As its microscopic justification  we consider an effective low energy theory \cite{BalentsPRB2021} described by the patch model considering only the V atoms of AV$_3$Sb$_5$, giving rise to vHS at the three \textbf{M} points in the Brillouin zone. We consider two-dimensional systems where topologically nontrivial configurations of the order parameter fields - vortices, can melt away the CDW order and restore the original lattice symmetry without destroying the quasiparticle gaps. We find that  there an interval of temperatures above the CDW phase transition where only  a composite order of the three CDWs can exist  while the individual CDW order parameters remain fluctuating. The latter ones condense at low temperatures.

We organize the rest of the paper as follows. In Section \ref{sec 1}, we present our working microscopic model which includes the interactions in the system, giving rise to the electronic CDW instability. Then, in Section \ref{sec 2}, we perform the mean field analysis of the CDW orders. In Section \ref{sec 3}, we consider  fluctuations of the CDW order parameters and present a GL free energy by incorporating the vortex configurations  by means  of dual fields. In Section \ref{sec 4}, we consider a simplified case, where only two CDW orders develop. We discuss appearance of the composite order in this model. Then in Section \ref{sec 5} we discuss the effects of doping away from the van Hove singularities. We argue that melting of the TrH phase in a doped system leads to emergence of a pseudogap regime resembling the one observed in the underdoped cuprates. At last we give a conclusion of our work in Section \ref{sec 6}.

\section{Model}\label{sec 1}
 The goal of our work is to describe fluctuations in the CDW regime of Kagome metals described by the patch model adopted by T. Park {\it et.al.} \cite{BalentsPRB2021}. A similar model leading to the same Ginzburg-Landau energy was used in \cite{MDennerbondPRL2021}. Both models consider just one vanadium orbital per site of the kagome lattice. The CDW order is believed to be electronically driven. However, there are some experiments which point to the role of phonons \cite{BOrtizPRM2021}. 

\begin{figure}[t]
\includegraphics[width=0.3 \textwidth]{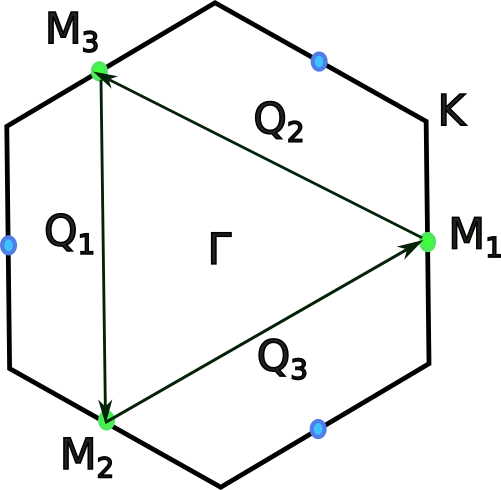}
\vspace{-0.2cm}
\caption{The hexagonal Brillouin zone, showing the high symmetry points. AV$_3$Sb$_5$ exhibit saddle points at the \textbf{M$_{1,2,3}$}, shown by green and blue circles. The \textbf{M$_a$} points are connected by the three nesting vectors \textbf{Q$_{1,2,3}$}, which are also the ordering wave-vectors of the CDW.}
\label{Fig:BZnesting}
\end{figure}

The first principles calculations \cite{ortiz2020cs,ZhaoDFT2021,hu2022rich} for the kagome metals AV$_3$Sb$_5$ show saddle points at the $\textbf{M}_a$ points of the hexagonal Brillouin zone, giving rise to the logarithimically divergent density of states. Hence we consider an effective low-energy model which takes into account only patches of the Fermi surface around the $\textbf{M}_a$ points in the Brillouin zone [ Fig. \ref{Fig:BZnesting}] of kagome metals AV$_3$Sb$_5$ and interactions among the fermionic states between these saddle points as was done in  \cite{BalentsPRB2021}.

The non-interacting Hamiltonian is given by 
 \begin{align}\label{eq:H_0}
 H_0 = \sum_{a=1}^3 \sum_{|k| < \Lambda}c^\dagger_{a\sigma}(k) [\epsilon_a(k) -\mu] c_{a\sigma}(k),
 \end{align}
where the single electron dispersion close to the saddle points $\textbf{M}_a$ are given by, 
 \begin{align}\label{eq:dispersion}
 \nonumber
 \epsilon_1 &= k_1(k_1+k_2),\\
 \nonumber
  \epsilon_2 &= -k_1k_2,\\
  \nonumber
   \epsilon_3 &= k_2(k_1+k_2),\\
 k_{1,2} &= k_x \pm \sqrt 3 k_y,
  \end{align}
 and $\mu$ is the chemical potential. Now, we consider electron-electron interactions among the fermions in the three patches close to the $M_a$ saddle points.
 \begin{align}\label{eq:hint}
 \nonumber
H_{int} &= \sum_{a\neq b}\sum_{k_1,k_2,k_3,k_4}\Big[ g_1(c^\dagger_{a, k_1,\sigma}c_{b,k_4,\sigma})(c^\dagger_{b,k_2,\sigma'}c_{a,k_3,\sigma'}) +\\
\nonumber
& g_2(c^\dagger_{a,k_1,\sigma}c_{a,k_4,\sigma})(c^\dagger_{b,k_2,\sigma'}c_{b,k_3,\sigma'}) +\\
\nonumber
& g_3(c^\dagger_{a,k_1,\sigma}c^\dagger_{a,k_2,-\sigma})(c_{b,k_3,-\sigma}c_{b,k_4,\sigma})\Big] \\
&+ \sum_{a}\sum_{k_1,k_2,k_3,k_4} g_4(c^\dagger_{ak_1,,\sigma}c_{a,k_4,\sigma})(c^\dagger_{a,k_2,-\sigma}c_{a,k_3,-\sigma}).
 \end{align}
 
The total effective Hamiltonian is given by
 \begin{align}\label{eq:Htot}
 H = H_0 + H_{int}    
 \end{align}
 In the Hamiltonian, $a = 1,2,3$ are the patch indices and $\vec{k}$ is momentum measured from the $\textbf{M}_a$. In the Eqn.(\ref{eq:hint}), we have the constraint $\vec{k}_1 + \vec{k}_2 + \vec{k}_3 + \vec{k}_4 = 0$. The coupling constants $g_1, g_2, g_3$ and $g_4$ represent interpatch exchange, interpatch density-density, Umklapp and intra-patch scattering terms respectively.
 
The parquet renormalization group (pRG) analysis \cite{BalentsPRB2021} suggests that an instability in the system occurs if the corresponding interaction strength becomes positive. For our work, we are interested only in various charge-density wave (CDW) instabilities. We do not consider \cite{BalentsPRB2021} interplay between the superconductivity and the CDW, as the superconductivity appears only at very low temperature. Now one can construct the following CDW type order parameters in the patch model. 

For the real CDW (rCDW), and imaginary CDW (iCDW), the order parameters are respectively,
\begin{align}
\Omega_{a} &\sim G_1 \sum_{k,\sigma} \langle c^\dagger_{a_{2}k}c_{a_{3} k^{\prime}}\rangle, \\
\Psi_{a} &\sim  \frac{G_2}{i} \sum_{k,\sigma} \langle c^\dagger_{a_{2}k}c_{a_{3} k^{\prime}}\rangle.
\end{align}
 The effective interaction strengths for the rCDW and iCDW are $G_1 = - 2g_1 + g_2  - g_3$ and $G_2 = - 2g_1 + g_2  + g_3$ respectively. Moreover, $\vec{k}^{\prime} = \vec{k} + \vec{Q}_{a}$. $\vec{Q}_{a}$ are the three nesting vectors connecting the $\textbf{M}_a$ and the ordering wave-vectors for the CDW order parameters as shown in the Fig. \ref{Fig:BZnesting}.
 
 The interactions are assumed to be quasi-local with $\Lambda$ being the UV cut-off. The Hamiltonian is $SU(2)\times Z_3\times U(1)$ invariant. 
 \\
  
 \section{Mean field theory}\label{sec 2}
 In this section we perform a mean-field decoupling of the Eqn.(\ref{eq:Htot}) in the CDW channels. This already suggests that we have $g_4 =0$, as the effective interactions for rCDW and iCDW do not depend on $g_4$. We keep both the rCDW and iCDW and derive a mean-field Hamiltonian and a GL free energy in terms of a complex CDW order parameters.
   
If the leading interaction term is $g_1$, one can perform the Hubbard-Stratonovich transformation,
 \begin{align}
 H_{mf} &=|V_{ab}|^2/2g_1 + \sum_{a>b} \Big[ V_{ab}c^+_{a\sigma}c_{b\sigma} + V^*_{ab}c^+_{b\sigma}c_{a\sigma}\Big] + \\
 \nonumber
 & \sum_{a=1}^3 \sum_{|k| < \Lambda} \epsilon_a(k) c^+_{a\sigma}(k)c_{a\sigma}(k). 
 \end{align}
 We introduce notations 
 $V_{12} = \Delta_3, ~~ V_{13} = \Delta_2, ~~ V_{23} = \Delta^*_{1}$.
Now, by integrating out the fermion field, we obtain the action in terms of the CDW order parameter fields  $\Delta_{a} =\Omega_{a} + i \Psi_{a} = |\Delta_{a}| e^{i\phi_{a}}$:
\begin{align}
F & =   \frac{1}{2g_1} \int d^2x d\tau |\Delta_{a}|^2 - \text{Tr} ~\text{ln} ~\mathcal{G}^{-1},
\end{align}
where $\mathcal{G}^{-1}$ is the inverse Green's function matrix. 
The electronic spectrum at the saddle point is determined by the equation: 
\begin{align}
 \Big|
 \begin{array}{ccc}
 -E + \epsilon_1 & \Delta_3 & \Delta_2\\
 \Delta^*_3 & -E + \epsilon_2 & \Delta_1^*\\
 \Delta_2^* & \Delta_1 & -E + \epsilon_3
 \end{array}
 \Big|=0,
\end{align}

\begin{figure}[t]
\includegraphics[width=0.4 \textwidth]{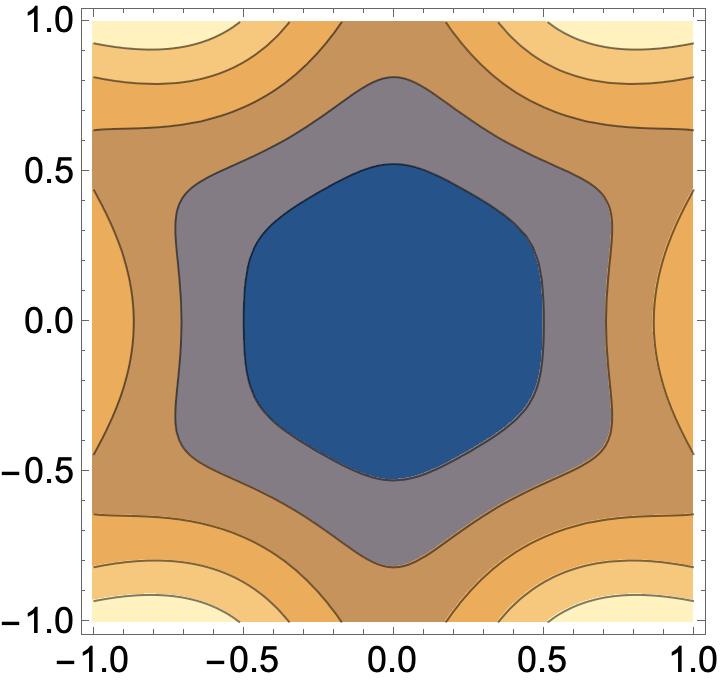}
\vspace{-0.2cm}
\caption{Contour plots of the gapless quasiparticle branch in the $-3Q$ phase. The Fermi surface is the boundary between the grey and the brown areas.}
\label{Fig:FS}
\vspace{-0.5cm}
\end{figure}

 The result is  
 \begin{align}\label{secular}
 \nonumber
 &(\epsilon_1-E)(\epsilon_2 -E)(\epsilon_3 -E)  +E(|\Delta_1|^2 + |\Delta_2|^2 +|\Delta_3|^2) \\
 \nonumber
 &-\epsilon_1|\Delta_1|^2 - \epsilon_2|\Delta_2|^2 - \epsilon_3|\Delta_3|^2 +\Delta_1\Delta_3\Delta_2 + \Delta_1^*\Delta_2^*\Delta^*_3 \\
 &=0. 
 \end{align}
 We assume that the fluctuations of moduli are gapped and consider the saddle point where  all $|\Delta_a|$ are equal. 
 It follows  from Eqn.(\ref{eq:dispersion}) that 
 \begin{align}
 \epsilon_1\epsilon_2 + \epsilon_1\epsilon_3 + \epsilon_2\epsilon_3 =0,
 \end{align}
 which leads to simplification of Eqn.(\ref{secular}) resulting in  
\begin{align}
\nonumber
 &(E\pm 1)^2(E\mp 2) - 3(k_x^2+k_y^2)(E^2-1) + 4k_x^2(k_x^2 -3k_y^2)^2\\
 &=0.
\end{align}
 where we set $|\Delta_a| =1$. According to \cite{NandkishorePRB2022}, plus sign in the first bracket  corresponds to the  $-3Q$ phase (SoD). In this phase there is a Fermi surface [see Fig. \ref{Fig:FS}]. The minus sign corresponds to the  $+3Q$ (TrH) phase where  the quasiparticle spectrum is fully gapped. 
At $g_3 \neq 0$, the mean-field spectrum should be corrected :
\begin{align}
\Delta_a \rightarrow \Delta_a + (g_3/g_1)\Delta_a^*,
\end{align}
This change does not modify the spectrum qualitatively though it modifies the Green's function.

\section{Ginzburg-Landau free energy}\label{sec 3}

We will follow the conclusions of the previous papers and, as we have mentioned above, consider the saddle point solution with all $|\Delta_a|$ being equal and treat fluctuations of the moduli of $\Delta$'s as gapped. Hence the subsequent  analysis of the GL free energy will include only phase fluctuations. In the absence of the Umklapp $g_3=0$ the only phase dependent term in the free energy density corresponds to the product of all three $\Delta$'s:
\begin{align}
\delta F = - G\cos(\phi_1+\phi_2+\phi_3).
\end{align}
Hence in the absence of the Umklapp  two phase fields remain critical in the low - temperature phase. However, if $g_3 \neq 0$ there is a contribution to the free energy density:
 \begin{align}
 g_3\Big[\Delta^2_a + (\Delta_a^*)^2] \sim g_3[\cos(2\phi_1) + \cos(2\phi_2) +\cos(2\phi_3)]
 \end{align}
 
\subsection{Fluctuations in the CDW order parameters}\label{ 3a}
 Now  we will consider phase fluctuations of the CDW order parameters which requires inclusion of the gradient terms. In two dimensions one must account for topologically nontrivial configurations of order parameter fields - vortices, which are being point-like objects with finite energy and can be thermally excited. To properly account for such configurations we regularize the model by putting it  on a suitable lattice with lattice constant $b$ and then taking a continuum limit.

The form of the free energy functional Eqn.(\ref{GL}) reflects the fact that the order parameters are periodic functions of $\phi_a$. This feature allows for topologically nontrivial configurations of the fields in the form of vortices - configurations where $\phi_a$ fields change by $2\pi$ along closed spacial loops. In the continuum limit such configurations are singular which explains the necessity for lattice regularization. It is well known that in 2D vortices can change a character of phase transitions. One way to take them into account in the continuous limit is to introduce  dual phase fields $\bar\phi_a$ \cite{JKKN1977}. In the present case is slightly unusual because in the region of interest the GL action contains the terms which depend on both $\phi$ and $\bar\phi$. The corresponding formalism was introduced in \cite{AleinerPRB2007} (see also \cite{tsvelik2014composite}). The  regularized GL free energy density is

 \begin{align} \label{GL}
 \nonumber
 {\cal F} /T&= -\frac{J}{T}\sum_a \sum_{<b>}\frac{1}{b^2}\cos\Big[\phi_a({\bf x}) - \phi_a({\bf x} + {\bf b})\Big]\\
 & - G\cos(\phi_1 +\phi_2 +\phi_3) + Ag_3\sum_a\cos(2\phi_a),
 \end{align}
where coefficient $A \sim \Delta^2/T >0$. 

Now we can follow the standard procedure and write down the the continuum limit of Eqn.(\ref{GL}) as  (in what follows  we will set the stiffness $J= 1$):
 \begin{align}\label{F}
 \nonumber
 {\cal F }/T&= \sum_a\Big[ \frac{1}{2T}(\partial_x\phi_a)^2 +\frac{T}{2}(\partial_x\bar\phi_a)^2 + i \partial_x\phi_a\partial_y\bar\phi_a \\
 & + Ag_3 \cos(2\phi_a) + \eta\cos(2\pi\bar\phi_a)\Big]- G\cos(\phi_1 +\phi_2 +\phi_3) . 
\end{align}
 The coupling $\eta$ is proportional to the vortex fugacity. The model Eqn.(\ref{F}) contains both original fields $\phi_a$ and their dual fields $\bar\phi_a$ which take care of the vortex configurations. The corresponding path integral for the partition function includes integration over both fields:
\begin{align}
Z = \int D\phi_a(x) D\bar\phi_a(x) \exp\Big(- \int d^2x {\cal F}/T\Big).
\end{align}

To determine whether the cosine terms are relevant or irrelevant, one has to  calculate their scaling dimensions. To compute the scaling dimensions of various perturbations, we start with the Gaussian model. The results are 
 
\begin{align}\label{ew:scaling1}
d_{g_3} = T/\pi, ~~ d_{\eta}= \pi/T, ~~ d_G = 3T/4\pi, 
 \end{align}

 The direct and dual operators cannot order simultaneously; this  creates an interesting situation at the transition where both of them are relevant. It is a nontrivial situation, see, for example \cite{tsvelik2014composite}.
 
In what follows we consider a limit of large $G$ when the sum of all phases is fixed \cite{WuPRL2021}. Now, we can make a transformation as follows:
 \begin{align}
 \phi_a &= \Phi/\sqrt 3 + (\sqrt{2/3})\vec e_a\vec\chi, \nonumber\\
 & {\bf e}_a = (1,0), ~(-1/2, \sqrt 3/2), ~(-1/2,-\sqrt 3/2).
 \end{align}
  with  $\vec\chi = (\chi_1,\chi_2)$ and treat $\Phi$ as gapped. 
 
In this case we get following the calculation shown in Appendix \ref{App A}, an effective free energy:
 \begin{align} \label{F2}
 \nonumber
 {\cal F}_{eff}/T &= \sum_{a=1}^3\Big[ \bar A g_3 \cos(\sqrt{8/3}{\bf e}_a\vec\chi) - B\cos(2\pi\sqrt{2}\vec\omega_a\bar{\vec\chi}) \Big]+ \\
 &\sum_{i=1,2}\Big[\frac{1}{2T}(\partial_x\chi_i)^2 +\frac{T}{2}(\partial_x\bar\chi_i)^2 +i \partial_x\chi_i\partial_y\bar\chi_i\Big], 
 \end{align}
 where  $\vec\omega_a = (0,1), (\sqrt 3, 1)/2, (\sqrt 3 ,-1)/2$,  $B\sim \eta^2$ and $\bar A = \langle\cos(\Phi/\sqrt 3)\rangle A$. 
\begin{figure}[t]
\includegraphics[width=0.48 \textwidth]{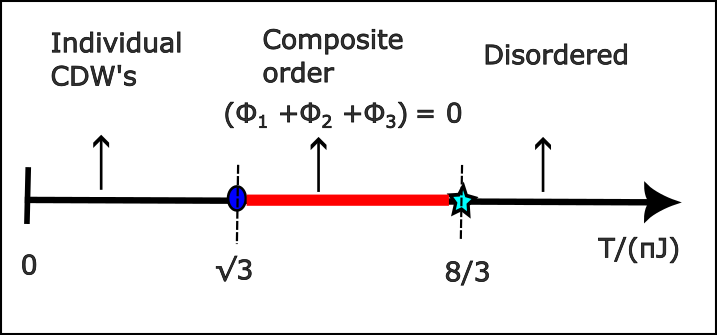}
\vspace{-0.2cm}
\caption{The phase diagram of the  kagome metal layer. For J =1, there is a crossover into a regime with 
 composite order around $\frac{T}{\pi} = \frac{8}{3}$, where the sum of the CDW phases are frozen. For a doping slightly away from the vHS ($\mu \neq 0$), it will exhibit a `pseudogap' -like behavior. Around $\frac{T}{\pi} = \sqrt{3}$ there is a  phase transition into the state where  individual CDW's order. For $g_3>0$ the low temperature phase breaks time-reversal symmetry. }
\label{Fig:Phase}
\end{figure} 

 The scaling dimensions of the cosines are
 \begin{align}
 d_{g_3} = 2T/3\pi, ~~ d_{B}  =2\pi/T
 \end{align}
The perturbations are relevant or irrelevant when the scaling dimension of the operators $d_{op} < D$ and $d_{op} > D$ respectively, $D$ being the spatial dimension of the system. We observe that below $T/\pi =3$ both direct and dual cosine terms are relevant provided the $G$-term is relevant which is true for $T/\pi <8/3$. 
 
 Below $T_{c1}/\pi = 8/3$, the $G$-term is relevant and the sum of all phases is frozen. However, above certain temperature $T_{c2}$ the vortices destroy the  order of individual CDWs. Only the product of their order parameters acquires a finite average, which we refer to as a composite order ($\Delta_1\Delta_2\Delta_3$) [see Fig. \ref{Fig:Phase}]. Since the periodicity of this order parameter coincides with the periodicity of the lattice, $T_{c1}$ is likely to mark a crossover. 

At $T_{c2}$ there is a  phase transition into a phase where individual phases are frozen which breaks the symmetry of lattice and may also break the time reversal symmetry (see below and also see Fig. \ref{Fig:Phase}).
  The character of the low temperature phase is determined by the  signs  of $G$ and $g_3$. At $G >0$ the product of $\Delta$'s have the same sign and we have the TrH order, at $G<0$ it is negative and we have the SoD pattern. If $g_3 <0$ the vacuum corresponds to $\chi_1 = \chi_2 =0$. This is $rCDW$ - real CDWs. If $g_3 >0$ there are degenerate vacua situated on a hexagonal lattice of $\chi_{1,2}$.  There are two inequivalent points $ \sqrt{8/3}(\chi_1,\chi_2) = (4\pi/3, 0)$ and $(2\pi/3, 2\pi/\sqrt 3)$. At each of these  vacua all $\Delta_a$'s are the same and are equal either to $\exp(4\pi i/3)$ or its complex conjugate. In the broken symmetry state one of this vacua is chosen which corresponds to complex $rCDW+iCDW$ with a broken time-reversal. This time-reversal symmetry breaking spontaneously induces orbital currents which can manifest in anomalous Hall effect \cite{yang2020giant}. The resulting real space pattern for the corresponding bond order can be either SoD or TrH along with the current order pattern as also discussed in \cite{BalentsPRB2021,MDennerbondPRL2021}.

  The  transition temperature is determined by the competition between normal and dual cosine perturbations. It can be estimated by comparing the mass scales generated by the competing operators. 
A relevant perturbation can drive a phase transition and the transition point can be determined by estimating the scale of mass gaps in the corresponding phases and comparing them with each other. The scale of the mass gap can be estimated by the fact that the contribution to the action of the relevant operator inducing a finite correlation length, becomes of the order of unity. Therefore, we notice that the phase transition from the composite order state to the state with individual CDWs occurs when  
  \begin{align}
  B^{1/(2-d_B)} \sim (\bar Ag_3)^{1/(2-d_{g_3})}.
  \end{align}
  Solving this equation with logarithmic accuracy we get the estimate for the transition temperature:
  \begin{align}
      T_{c2}/\pi = \frac{3}{2}\Big(1- \alpha +\sqrt{1- 2\alpha/3+\alpha^2}\Big), ~~\alpha = \frac{\ln(\bar Ag_3)}{\ln B}.
  \end{align}
  For comparable coupling constants it yields $T_{c2}/\pi = \sqrt 3$ and $d^* = 2/\sqrt 3$. 
  
The model Eqn.(\ref{F2}) belongs to the class of affine XY models which have been studied in connection to the problem of quark confinement \cite{anber20122d,anber20133}.  Although this particular model has not been studied, some insights can be drawn. An affine XY model with different operators was studied numerically in \cite{anber20133}  and the results indicate that the transition  is probably weak first order. The hysteresis, however, has not been observed which leaves a possibility of a second order phase transition. The uncertainty remains and first order transition remains  a possibility also for our case. If, however, it is a second order transition, then following the results for another similar model \cite{Philippe}, we suggest  that it would belong to the $\mathbb Z_3$ Potts universality class \cite{wu1982potts}. This suggest that  the  critical exponents  are $\nu = \frac{5}{6}$, $\eta = \frac{4}{15}$ \cite{francesco2012conformal}.
 
\section{Simplified case}\label{sec 4}
 
The purpose of this section is to study an example of a treatable model describing a phase transition driven by  mutually dual cosines. This model describes the case when only two CDWs develop:  $\Delta_3 =0, ~|\Delta_1| = |\Delta_2|$, describing a nematicity \cite{grandi2023theory} in the system. Then there are two  phases $\phi_1$ and $\phi_2$, whose fluctuations are described by action Eqn.(\ref{GL}). Once there sum is frozen we arrive at model (Eqn.(\ref{F2})) with a single pair of fields $\chi, \bar\chi$. This situation was studied in \cite{tsvelik2014composite} repeat the calculations here for illustrative purposes. 

 In this case at $T/\pi = 1$ the scaling dimensions of the cosines are equal to 1, giving rise to comparable values of $\bar Ag_3$ and $B$. At this point, the bosonic action (Eqn.(\ref{F2})) can be refermionized and recast as a model of relativistic fermions with two kinds of mass terms \cite{tsvelik2012parafermion}: 
\begin{align}
\nonumber
 {\cal F}_{eff}/T &=  R^+(\partial_y -i\partial_x) R +  L^+(\partial_y +i\partial_x )L + \nonumber\\
 & \bar Ag_3 (R^+L + L^+R) 
  + B(RL + L^+R^+).
 \end{align}
 The next step is two express the Dirac fermions in terms of Majoranas: 
 \begin{align}
R,L = \frac{1}{\sqrt{2}}\Big[\rho_{R,L}^{(+)} + i \rho_{R,L}^{(-)}\Big].
\end{align}
 As a result we get two separate models for Majorana fermions with masses 
$m_{\pm} = \bar Ag_3 \pm B$. Each Majorana species corresponds to 2D Ising model where the mass is proportional  to $(T-T_c)$. For any sign of $g_3$ the transition occurs only for one Majorana species. As was shown in  \cite{tsvelik2014composite} the CDW order parameter (for instance, $\Delta_1$, since in the given case $\Delta_2 = \Delta_1^*$) can be written as 
 \begin{align}
 \Delta = i\sigma_+\sigma_- + \mu_+\mu_-,
 \end{align}
 where $\sigma_{\pm}$ are the order and $\mu_{\pm}$ are the disorder parameters of the Ising models with masses $m_{\pm}$. One of these models is always in the ordered $\langle\sigma\rangle\neq 0$ or disordered $(\langle\mu\rangle \neq 0)$ phase, the other one undergoes a phase transition. It can be shown that in the part of the phase diagram where $m_{\pm} > 0$, the expectation value of $\sigma_{1/2}$ is non vanishing, whereas average of $\mu_{1/2}$ vanishes. Also, for $m_{\pm} < 0$, the average of $\sigma_{\pm}$ vanishes, while average of $\mu_{\pm}$ becomes finite. 

  The $\mathbb Z_3$ symmetric case is more complicated. If the transition is of the second order then some insight can be drawn from  \cite{Philippe}, where a similar model at the transition point was represented as a sum of the critical $\mathbb Z_3$ Potts model and a W$_3$ Conformal Field Theory perturbed by a relevant operator:
  \begin{align}
  H = H_{\mathbb Z_3}^0 + H_{W_3}^0 - \gamma \Phi_{\lambda_1 +\lambda_2,0},
  \end{align} 
  where $\lambda$'s are fundamental weights of the SU(3) group. The perturbed $W_3$  theory is massive.
  
  \section{Doping}\label{sec 5}
  
All previous calculations remain valid if the chemical potential is slightly away from the vHS. In the  low temperature state the CDW order will lead to reconstruction of the Fermi surface through Brillouin zone folding with appearance  of small Fermi pockets as described in, for example, in \cite{zhou2022chern}. Once the temperature exceeds $T_{c2}$ the individual CDW order will melt, but the spectral gaps will survive. The system will enter in a pseudogap regime similar to the one observed in the underdoped cuprates where below the certain crossover temperature most of the original Fermi surface gradually fades away  and the low energy spectral weight is concentrated  at small pockets. As in the cuprates the predicted  crossover is not accompanied by a broken lattice symmetry. The ideas that melting of the low temperature Neel order may explain the observations of the Fermi surface arcs.\cite{Sachdev2014fermi}

 \section{Conclusions}\label{sec 6}
 In this work we have studied a fluctuation regime in the CDW order, within an effective low energy interacting patch model \cite{BalentsPRB2021} describing a layered kagome system or a two-dimensional film. We study the fluctuation by considering a field theoretic technique which allows us to treat simultaneously the effects of  the discrete symmetry breaking order and the vortex physics. 
 We observe that the interplay of fluctuations and topology (vortices) {\it in two dimensions} leads to  formation of a special regime where the individual  low temperature CDW orders melt restoring the lattice symmetry but keeping intact the quasiparticle gaps. At further lowering of the temperature the system undergoes a phase transition into the phase with individual CDW order. 
 
The suggested mechanism  is similar to the mechanism of formation  of charge $6e$ superconducting condensate described theoretically in \cite{Agterberg2008,Berg2009,Asvin2009,zhou2022chern} and recently observed in the thin flakes of the kagome superconductor CsV$_3$Sb$_5$ \cite{6e2023}. The measurements were performed on mesoscopic CsV$_3$Sb$_5$ rings are fabricated by etching the kagome
superconductor thin flakes exfoliated from bulk samples. We suggest a similar arrangement for the CDW experiments. 
 We identify the CDW transition as  belonging to the $\mathbb Z_3$ Potts universality class. 
 
\section{Acknowledgements}

 We are grateful to Andrey Chubukov, Philippe Lecheminant and Dmitry Kovrizhin  for valuable discussions. We are grateful to Philippe Lecheminant for attracting our attention to several  beautiful papers relevant to the present topic.  This work was supported by Office of Basic Energy Sciences, Material
 Sciences and Engineering Division, U.S. Department of Energy (DOE)
 under Contracts No. DE-SC0012704.
 
\appendix
\section{Free energy for a large value of G:}\label{App A}
The free energy considered in the section \ref{ 3a} is given by:
\begin{align}
 \nonumber
 {\cal F }/T&= \sum_a\Big[ \frac{1}{2T}(\partial_x\phi_a)^2 +\frac{T}{2}(\partial_x\bar\phi_a)^2 + i \partial_x\phi_a\partial_y\bar\phi_a \\
 & + Ag_3 \cos(2\phi_a) + \eta\cos(2\pi\bar\phi_a)\Big]- G\cos(\phi_1 +\phi_2 +\phi_3) . 
\end{align}

At $\eta =0$ one can integrate over $\partial_x\bar\phi_a$:
\begin{align}
\nonumber
 &\int D\bar\phi_a \exp\{-\int d^2x [ \frac{T}{2}(\partial_x\bar\phi_a)^2 + i \partial_x\phi_a\partial_y\bar\phi_a ] \}\\& = const.\int D\bar\phi_a \exp\{ - \int d^2x [ \frac{T}{2}(\partial_x\bar\phi_a)^2 - i \partial_y\phi_a\partial_x\bar\phi_a ] \} \nonumber\\
& \sim \exp[-\frac{1}{2T}\int d^2 x (\partial_y\phi)^2]
\end{align}
The result is the partition function for $\phi$ fields:
\begin{align}
\nonumber
Z[\phi] &= \int D\phi_a \exp\{- \int d^2x[ \frac{1}{2T}[(\partial_x\phi_a)^2 + \\
&(\partial_y\phi_a)^2] + Ag_3\cos(2\phi_a) - G\cos(\phi_1 +\phi_2 +\phi_3)\}.
\end{align}
In a similar way at $g_3 =0$ we can integrate out the $\phi$-fields.

In the limit of large G, we can consider the sum of all phases to be fixed. Hence the GL free energy can be transformed with
 \begin{align}
 \nonumber
 \phi_1 &= \Phi/\sqrt 3 + (\sqrt{2/3})\chi_{1}, \\
 \nonumber
 \phi_2 &= \Phi/\sqrt 3 + (\sqrt{2/3})(-\chi_{1}/2 + \sqrt{3}/2 \chi_2), \\
 \phi_3 &= \Phi/\sqrt 3 - (\sqrt{2/3})(\chi_{1}/2 + \sqrt{3}/2 \chi_2).
 \end{align}
In this case, $\cos(\phi_1 + \phi_2 + \phi_3) = \cos(\sqrt{3}\Phi)$. 
When $\Phi$ is frozen, the dual field $\bar\Phi = \sum_a\bar\phi_{a}$ fluctuates strongly so that  correlators of the dual exponents decay exponentially.  Then the dual perturbation is generated in the second order in $\eta$:
\begin{align}
    &\eta^2\int d^2 x \exp\{i\bar\Phi/\sqrt 3 + \sqrt{8/3\pi}\vec e_a\bar{\vec\chi}\}_{\bf r}\times \nonumber\\
    &\exp\{-i\bar\Phi/\sqrt 3 - \sqrt{8/3\pi}\vec e_b\bar{\vec\chi}\}_{{\bf r}+{\bf x}},  
\end{align}
giving rise to the operators
 \begin{align}
 \cos[\sqrt{8/3}\pi (\vec e_a - \vec e_b)\bar{\vec\chi}],
 \end{align}
 with scaling dimension $2\pi/T$.  
In that case we get 
 \begin{align} \label{F21}
 \nonumber
 {\cal F}_{eff}/T &= \sum_{a=1}^3\Big[ \bar A g_3 \cos(\sqrt{8/3}{\bf e}_a\vec\chi) - B\cos(2\pi\sqrt{2}{\vec\omega_a}\bar{\vec\chi}) \Big]+ \\
 &\sum_{i=1,2}\Big[\frac{1}{2T}(\partial_x\chi_i)^2 +\frac{T}{2}(\partial_x\bar\chi_i)^2 +i \partial_x\chi_i\partial_y\bar\chi_i\Big].
 \end{align}
 with $\vec\omega_a = (0,1), (\sqrt 3, 1)/2, (\sqrt 3 ,-1)/2$. 
 More explicitly: 
 \begin{align}
 \nonumber
&\sum_a\cos(\sqrt{8/3}{\bf e}_a\vec\chi) = \cos[\sqrt{8/3}\chi_1] + \cos[\sqrt{2/3}(\chi_1 -\sqrt 3\chi_2)] +\\
&\cos[\sqrt{2/3}(\chi_1 +\sqrt 3\chi_2)] ,
 \end{align}
 and,
 \begin{align}
 \nonumber
 \sum_a \cos(2\pi\sqrt{2}\vec\omega_a\bar{\vec\chi}) \Big]& = \cos[2\pi\sqrt 2\bar\chi_2] +\cos[\pi\sqrt 2(\bar\chi_2 +\sqrt 3\bar\chi_1)] \\
 &+\cos[\pi\sqrt 2(\bar\chi_2 -\sqrt 3\bar\chi_1)] .
 \end{align}

\bibliographystyle{apsrev4-1}
\bibliography{Kagome}

\begin{thebibliography}{82}%
\makeatletter
\providecommand \@ifxundefined [1]{%
 \@ifx{#1\undefined}
}%
\providecommand \@ifnum [1]{%
 \ifnum #1\expandafter \@firstoftwo
 \else \expandafter \@secondoftwo
 \fi
}%
\providecommand \@ifx [1]{%
 \ifx #1\expandafter \@firstoftwo
 \else \expandafter \@secondoftwo
 \fi
}%
\providecommand \natexlab [1]{#1}%
\providecommand \enquote  [1]{``#1''}%
\providecommand \bibnamefont  [1]{#1}%
\providecommand \bibfnamefont [1]{#1}%
\providecommand \citenamefont [1]{#1}%
\providecommand \href@noop [0]{\@secondoftwo}%
\providecommand \href [0]{\begingroup \@sanitize@url \@href}%
\providecommand \@href[1]{\@@startlink{#1}\@@href}%
\providecommand \@@href[1]{\endgroup#1\@@endlink}%
\providecommand \@sanitize@url [0]{\catcode `\\12\catcode `\$12\catcode
  `\&12\catcode `\#12\catcode `\^12\catcode `\_12\catcode `\%12\relax}%
\providecommand \@@startlink[1]{}%
\providecommand \@@endlink[0]{}%
\providecommand \url  [0]{\begingroup\@sanitize@url \@url }%
\providecommand \@url [1]{\endgroup\@href {#1}{\urlprefix }}%
\providecommand \urlprefix  [0]{URL }%
\providecommand \Eprint [0]{\href }%
\providecommand \doibase [0]{http://dx.doi.org/}%
\providecommand \selectlanguage [0]{\@gobble}%
\providecommand \bibinfo  [0]{\@secondoftwo}%
\providecommand \bibfield  [0]{\@secondoftwo}%
\providecommand \translation [1]{[#1]}%
\providecommand \BibitemOpen [0]{}%
\providecommand \bibitemStop [0]{}%
\providecommand \bibitemNoStop [0]{.\EOS\space}%
\providecommand \EOS [0]{\spacefactor3000\relax}%
\providecommand \BibitemShut  [1]{\csname bibitem#1\endcsname}%
\let\auto@bib@innerbib\@empty
\bibitem [{\citenamefont {Kennes}\ \emph {et~al.}(2021)\citenamefont {Kennes},
  \citenamefont {Claassen}, \citenamefont {Xian}, \citenamefont {Georges},
  \citenamefont {Millis}, \citenamefont {Hone}, \citenamefont {Dean},
  \citenamefont {Basov}, \citenamefont {Pasupathy},\ and\ \citenamefont
  {Rubio}}]{kennes2021moire}%
  \BibitemOpen
  \bibfield  {author} {\bibinfo {author} {\bibfnamefont {D.~M.}\ \bibnamefont
  {Kennes}}, \bibinfo {author} {\bibfnamefont {M.}~\bibnamefont {Claassen}},
  \bibinfo {author} {\bibfnamefont {L.}~\bibnamefont {Xian}}, \bibinfo {author}
  {\bibfnamefont {A.}~\bibnamefont {Georges}}, \bibinfo {author} {\bibfnamefont
  {A.~J.}\ \bibnamefont {Millis}}, \bibinfo {author} {\bibfnamefont
  {J.}~\bibnamefont {Hone}}, \bibinfo {author} {\bibfnamefont {C.~R.}\
  \bibnamefont {Dean}}, \bibinfo {author} {\bibfnamefont {D.}~\bibnamefont
  {Basov}}, \bibinfo {author} {\bibfnamefont {A.~N.}\ \bibnamefont
  {Pasupathy}}, \ and\ \bibinfo {author} {\bibfnamefont {A.}~\bibnamefont
  {Rubio}},\ }\href {https://www.nature.com/articles/s41567-020-01154-3}
  {\bibfield  {journal} {\bibinfo  {journal} {Nature Physics}\ }\textbf
  {\bibinfo {volume} {17}},\ \bibinfo {pages} {155} (\bibinfo {year}
  {2021})}\BibitemShut {NoStop}%
\bibitem [{\citenamefont {Dzero}\ \emph {et~al.}(2016)\citenamefont {Dzero},
  \citenamefont {Xia}, \citenamefont {Galitski},\ and\ \citenamefont
  {Coleman}}]{dzero2016topological}%
  \BibitemOpen
  \bibfield  {author} {\bibinfo {author} {\bibfnamefont {M.}~\bibnamefont
  {Dzero}}, \bibinfo {author} {\bibfnamefont {J.}~\bibnamefont {Xia}}, \bibinfo
  {author} {\bibfnamefont {V.}~\bibnamefont {Galitski}}, \ and\ \bibinfo
  {author} {\bibfnamefont {P.}~\bibnamefont {Coleman}},\ }\href
  {https://doi.org/10.1146/annurev-conmatphys-031214-014749} {\bibfield
  {journal} {\bibinfo  {journal} {Annual Review of Condensed Matter Physics}\
  }\textbf {\bibinfo {volume} {7}},\ \bibinfo {pages} {249} (\bibinfo {year}
  {2016})}\BibitemShut {NoStop}%
\bibitem [{\citenamefont {Ortiz}\ \emph {et~al.}(2019)\citenamefont {Ortiz},
  \citenamefont {Gomes}, \citenamefont {Morey}, \citenamefont {Winiarski},
  \citenamefont {Bordelon}, \citenamefont {Mangum}, \citenamefont {Oswald},
  \citenamefont {Rodriguez-Rivera}, \citenamefont {Neilson}, \citenamefont
  {Wilson}, \citenamefont {Ertekin}, \citenamefont {McQueen},\ and\
  \citenamefont {Toberer}}]{BOrtizPRM2019}%
  \BibitemOpen
  \bibfield  {author} {\bibinfo {author} {\bibfnamefont {B.~R.}\ \bibnamefont
  {Ortiz}}, \bibinfo {author} {\bibfnamefont {L.~C.}\ \bibnamefont {Gomes}},
  \bibinfo {author} {\bibfnamefont {J.~R.}\ \bibnamefont {Morey}}, \bibinfo
  {author} {\bibfnamefont {M.}~\bibnamefont {Winiarski}}, \bibinfo {author}
  {\bibfnamefont {M.}~\bibnamefont {Bordelon}}, \bibinfo {author}
  {\bibfnamefont {J.~S.}\ \bibnamefont {Mangum}}, \bibinfo {author}
  {\bibfnamefont {I.~W.~H.}\ \bibnamefont {Oswald}}, \bibinfo {author}
  {\bibfnamefont {J.~A.}\ \bibnamefont {Rodriguez-Rivera}}, \bibinfo {author}
  {\bibfnamefont {J.~R.}\ \bibnamefont {Neilson}}, \bibinfo {author}
  {\bibfnamefont {S.~D.}\ \bibnamefont {Wilson}}, \bibinfo {author}
  {\bibfnamefont {E.}~\bibnamefont {Ertekin}}, \bibinfo {author} {\bibfnamefont
  {T.~M.}\ \bibnamefont {McQueen}}, \ and\ \bibinfo {author} {\bibfnamefont
  {E.~S.}\ \bibnamefont {Toberer}},\ }\href {\doibase
  10.1103/PhysRevMaterials.3.094407} {\bibfield  {journal} {\bibinfo  {journal}
  {Phys. Rev. Mater.}\ }\textbf {\bibinfo {volume} {3}},\ \bibinfo {pages}
  {094407} (\bibinfo {year} {2019})}\BibitemShut {NoStop}%
\bibitem [{\citenamefont {Ortiz}\ \emph {et~al.}(2020)\citenamefont {Ortiz},
  \citenamefont {Teicher}, \citenamefont {Hu}, \citenamefont {Zuo},
  \citenamefont {Sarte}, \citenamefont {Schueller}, \citenamefont {Abeykoon},
  \citenamefont {Krogstad}, \citenamefont {Rosenkranz}, \citenamefont {Osborn},
  \citenamefont {Seshadri}, \citenamefont {Balents}, \citenamefont {He},\ and\
  \citenamefont {Wilson}}]{ortiz2020cs}%
  \BibitemOpen
  \bibfield  {author} {\bibinfo {author} {\bibfnamefont {B.~R.}\ \bibnamefont
  {Ortiz}}, \bibinfo {author} {\bibfnamefont {S.~M.~L.}\ \bibnamefont
  {Teicher}}, \bibinfo {author} {\bibfnamefont {Y.}~\bibnamefont {Hu}},
  \bibinfo {author} {\bibfnamefont {J.~L.}\ \bibnamefont {Zuo}}, \bibinfo
  {author} {\bibfnamefont {P.~M.}\ \bibnamefont {Sarte}}, \bibinfo {author}
  {\bibfnamefont {E.~C.}\ \bibnamefont {Schueller}}, \bibinfo {author}
  {\bibfnamefont {A.~M.~M.}\ \bibnamefont {Abeykoon}}, \bibinfo {author}
  {\bibfnamefont {M.~J.}\ \bibnamefont {Krogstad}}, \bibinfo {author}
  {\bibfnamefont {S.}~\bibnamefont {Rosenkranz}}, \bibinfo {author}
  {\bibfnamefont {R.}~\bibnamefont {Osborn}}, \bibinfo {author} {\bibfnamefont
  {R.}~\bibnamefont {Seshadri}}, \bibinfo {author} {\bibfnamefont
  {L.}~\bibnamefont {Balents}}, \bibinfo {author} {\bibfnamefont
  {J.}~\bibnamefont {He}}, \ and\ \bibinfo {author} {\bibfnamefont {S.~D.}\
  \bibnamefont {Wilson}},\ }\href {\doibase 10.1103/PhysRevLett.125.247002}
  {\bibfield  {journal} {\bibinfo  {journal} {Phys. Rev. Lett.}\ }\textbf
  {\bibinfo {volume} {125}},\ \bibinfo {pages} {247002} (\bibinfo {year}
  {2020})}\BibitemShut {NoStop}%
\bibitem [{\citenamefont {Ortiz}\ \emph
  {et~al.}(2021{\natexlab{a}})\citenamefont {Ortiz}, \citenamefont {Sarte},
  \citenamefont {Kenney}, \citenamefont {Graf}, \citenamefont {Teicher},
  \citenamefont {Seshadri},\ and\ \citenamefont {Wilson}}]{Ortiztopometal2021}%
  \BibitemOpen
  \bibfield  {author} {\bibinfo {author} {\bibfnamefont {B.~R.}\ \bibnamefont
  {Ortiz}}, \bibinfo {author} {\bibfnamefont {P.~M.}\ \bibnamefont {Sarte}},
  \bibinfo {author} {\bibfnamefont {E.~M.}\ \bibnamefont {Kenney}}, \bibinfo
  {author} {\bibfnamefont {M.~J.}\ \bibnamefont {Graf}}, \bibinfo {author}
  {\bibfnamefont {S.~M.~L.}\ \bibnamefont {Teicher}}, \bibinfo {author}
  {\bibfnamefont {R.}~\bibnamefont {Seshadri}}, \ and\ \bibinfo {author}
  {\bibfnamefont {S.~D.}\ \bibnamefont {Wilson}},\ }\href {\doibase
  10.1103/PhysRevMaterials.5.034801} {\bibfield  {journal} {\bibinfo  {journal}
  {Phys. Rev. Mater.}\ }\textbf {\bibinfo {volume} {5}},\ \bibinfo {pages}
  {034801} (\bibinfo {year} {2021}{\natexlab{a}})}\BibitemShut {NoStop}%
\bibitem [{\citenamefont {Jiang}\ \emph {et~al.}(2021)\citenamefont {Jiang},
  \citenamefont {Yin}, \citenamefont {Denner}, \citenamefont {Shumiya},
  \citenamefont {Ortiz}, \citenamefont {Xu}, \citenamefont {Guguchia},
  \citenamefont {He}, \citenamefont {Hossain}, \citenamefont {Liu} \emph
  {et~al.}}]{jiang2021unconventional}%
  \BibitemOpen
  \bibfield  {author} {\bibinfo {author} {\bibfnamefont {Y.-X.}\ \bibnamefont
  {Jiang}}, \bibinfo {author} {\bibfnamefont {J.-X.}\ \bibnamefont {Yin}},
  \bibinfo {author} {\bibfnamefont {M.~M.}\ \bibnamefont {Denner}}, \bibinfo
  {author} {\bibfnamefont {N.}~\bibnamefont {Shumiya}}, \bibinfo {author}
  {\bibfnamefont {B.~R.}\ \bibnamefont {Ortiz}}, \bibinfo {author}
  {\bibfnamefont {G.}~\bibnamefont {Xu}}, \bibinfo {author} {\bibfnamefont
  {Z.}~\bibnamefont {Guguchia}}, \bibinfo {author} {\bibfnamefont
  {J.}~\bibnamefont {He}}, \bibinfo {author} {\bibfnamefont {M.~S.}\
  \bibnamefont {Hossain}}, \bibinfo {author} {\bibfnamefont {X.}~\bibnamefont
  {Liu}},  \emph {et~al.},\ }\href
  {https://www.nature.com/articles/s41563-021-01034-y} {\bibfield  {journal}
  {\bibinfo  {journal} {Nature materials}\ }\textbf {\bibinfo {volume} {20}},\
  \bibinfo {pages} {1353} (\bibinfo {year} {2021})}\BibitemShut {NoStop}%
\bibitem [{\citenamefont {Li}\ \emph {et~al.}(2021)\citenamefont {Li},
  \citenamefont {Zhang}, \citenamefont {Yilmaz}, \citenamefont {Pai},
  \citenamefont {Marvinney}, \citenamefont {Said}, \citenamefont {Yin},
  \citenamefont {Gong}, \citenamefont {Tu}, \citenamefont {Vescovo},
  \citenamefont {Nelson}, \citenamefont {Moore}, \citenamefont {Murakami},
  \citenamefont {Lei}, \citenamefont {Lee}, \citenamefont {Lawrie},\ and\
  \citenamefont {Miao}}]{LiPRXphonon2021}%
  \BibitemOpen
  \bibfield  {author} {\bibinfo {author} {\bibfnamefont {H.}~\bibnamefont
  {Li}}, \bibinfo {author} {\bibfnamefont {T.~T.}\ \bibnamefont {Zhang}},
  \bibinfo {author} {\bibfnamefont {T.}~\bibnamefont {Yilmaz}}, \bibinfo
  {author} {\bibfnamefont {Y.~Y.}\ \bibnamefont {Pai}}, \bibinfo {author}
  {\bibfnamefont {C.~E.}\ \bibnamefont {Marvinney}}, \bibinfo {author}
  {\bibfnamefont {A.}~\bibnamefont {Said}}, \bibinfo {author} {\bibfnamefont
  {Q.~W.}\ \bibnamefont {Yin}}, \bibinfo {author} {\bibfnamefont {C.~S.}\
  \bibnamefont {Gong}}, \bibinfo {author} {\bibfnamefont {Z.~J.}\ \bibnamefont
  {Tu}}, \bibinfo {author} {\bibfnamefont {E.}~\bibnamefont {Vescovo}},
  \bibinfo {author} {\bibfnamefont {C.~S.}\ \bibnamefont {Nelson}}, \bibinfo
  {author} {\bibfnamefont {R.~G.}\ \bibnamefont {Moore}}, \bibinfo {author}
  {\bibfnamefont {S.}~\bibnamefont {Murakami}}, \bibinfo {author}
  {\bibfnamefont {H.~C.}\ \bibnamefont {Lei}}, \bibinfo {author} {\bibfnamefont
  {H.~N.}\ \bibnamefont {Lee}}, \bibinfo {author} {\bibfnamefont {B.~J.}\
  \bibnamefont {Lawrie}}, \ and\ \bibinfo {author} {\bibfnamefont
  {H.}~\bibnamefont {Miao}},\ }\href {\doibase 10.1103/PhysRevX.11.031050}
  {\bibfield  {journal} {\bibinfo  {journal} {Phys. Rev. X}\ }\textbf {\bibinfo
  {volume} {11}},\ \bibinfo {pages} {031050} (\bibinfo {year}
  {2021})}\BibitemShut {NoStop}%
\bibitem [{\citenamefont {Uykur}\ \emph {et~al.}(2022)\citenamefont {Uykur},
  \citenamefont {Ortiz}, \citenamefont {Wilson}, \citenamefont {Dressel},\ and\
  \citenamefont {Tsirlin}}]{uykur2022optical}%
  \BibitemOpen
  \bibfield  {author} {\bibinfo {author} {\bibfnamefont {E.}~\bibnamefont
  {Uykur}}, \bibinfo {author} {\bibfnamefont {B.~R.}\ \bibnamefont {Ortiz}},
  \bibinfo {author} {\bibfnamefont {S.~D.}\ \bibnamefont {Wilson}}, \bibinfo
  {author} {\bibfnamefont {M.}~\bibnamefont {Dressel}}, \ and\ \bibinfo
  {author} {\bibfnamefont {A.~A.}\ \bibnamefont {Tsirlin}},\ }\href
  {https://www.nature.com/articles/s41535-021-00420-8} {\bibfield  {journal}
  {\bibinfo  {journal} {npj Quantum Materials}\ }\textbf {\bibinfo {volume}
  {7}},\ \bibinfo {pages} {16} (\bibinfo {year} {2022})}\BibitemShut {NoStop}%
\bibitem [{\citenamefont {Ortiz}\ \emph
  {et~al.}(2021{\natexlab{b}})\citenamefont {Ortiz}, \citenamefont {Teicher},
  \citenamefont {Kautzsch}, \citenamefont {Sarte}, \citenamefont {Ratcliff},
  \citenamefont {Harter}, \citenamefont {Ruff}, \citenamefont {Seshadri},\ and\
  \citenamefont {Wilson}}]{OrtizCDWPRX2021}%
  \BibitemOpen
  \bibfield  {author} {\bibinfo {author} {\bibfnamefont {B.~R.}\ \bibnamefont
  {Ortiz}}, \bibinfo {author} {\bibfnamefont {S.~M.~L.}\ \bibnamefont
  {Teicher}}, \bibinfo {author} {\bibfnamefont {L.}~\bibnamefont {Kautzsch}},
  \bibinfo {author} {\bibfnamefont {P.~M.}\ \bibnamefont {Sarte}}, \bibinfo
  {author} {\bibfnamefont {N.}~\bibnamefont {Ratcliff}}, \bibinfo {author}
  {\bibfnamefont {J.}~\bibnamefont {Harter}}, \bibinfo {author} {\bibfnamefont
  {J.~P.~C.}\ \bibnamefont {Ruff}}, \bibinfo {author} {\bibfnamefont
  {R.}~\bibnamefont {Seshadri}}, \ and\ \bibinfo {author} {\bibfnamefont
  {S.~D.}\ \bibnamefont {Wilson}},\ }\href {\doibase
  10.1103/PhysRevX.11.041030} {\bibfield  {journal} {\bibinfo  {journal} {Phys.
  Rev. X}\ }\textbf {\bibinfo {volume} {11}},\ \bibinfo {pages} {041030}
  (\bibinfo {year} {2021}{\natexlab{b}})}\BibitemShut {NoStop}%
\bibitem [{\citenamefont {Tan}\ \emph {et~al.}(2021)\citenamefont {Tan},
  \citenamefont {Liu}, \citenamefont {Wang},\ and\ \citenamefont
  {Yan}}]{TanPRL2021}%
  \BibitemOpen
  \bibfield  {author} {\bibinfo {author} {\bibfnamefont {H.}~\bibnamefont
  {Tan}}, \bibinfo {author} {\bibfnamefont {Y.}~\bibnamefont {Liu}}, \bibinfo
  {author} {\bibfnamefont {Z.}~\bibnamefont {Wang}}, \ and\ \bibinfo {author}
  {\bibfnamefont {B.}~\bibnamefont {Yan}},\ }\href {\doibase
  10.1103/PhysRevLett.127.046401} {\bibfield  {journal} {\bibinfo  {journal}
  {Phys. Rev. Lett.}\ }\textbf {\bibinfo {volume} {127}},\ \bibinfo {pages}
  {046401} (\bibinfo {year} {2021})}\BibitemShut {NoStop}%
\bibitem [{\citenamefont {Denner}\ \emph {et~al.}(2021)\citenamefont {Denner},
  \citenamefont {Thomale},\ and\ \citenamefont {Neupert}}]{MDennerbondPRL2021}%
  \BibitemOpen
  \bibfield  {author} {\bibinfo {author} {\bibfnamefont {M.~M.}\ \bibnamefont
  {Denner}}, \bibinfo {author} {\bibfnamefont {R.}~\bibnamefont {Thomale}}, \
  and\ \bibinfo {author} {\bibfnamefont {T.}~\bibnamefont {Neupert}},\ }\href
  {\doibase 10.1103/PhysRevLett.127.217601} {\bibfield  {journal} {\bibinfo
  {journal} {Phys. Rev. Lett.}\ }\textbf {\bibinfo {volume} {127}},\ \bibinfo
  {pages} {217601} (\bibinfo {year} {2021})}\BibitemShut {NoStop}%
\bibitem [{\citenamefont {Feng}\ \emph {et~al.}(2021)\citenamefont {Feng},
  \citenamefont {Jiang}, \citenamefont {Wang},\ and\ \citenamefont
  {Hu}}]{feng2021chiral}%
  \BibitemOpen
  \bibfield  {author} {\bibinfo {author} {\bibfnamefont {X.}~\bibnamefont
  {Feng}}, \bibinfo {author} {\bibfnamefont {K.}~\bibnamefont {Jiang}},
  \bibinfo {author} {\bibfnamefont {Z.}~\bibnamefont {Wang}}, \ and\ \bibinfo
  {author} {\bibfnamefont {J.}~\bibnamefont {Hu}},\ }\href {\doibase
  https://doi.org/10.1016/j.scib.2021.04.043} {\bibfield  {journal} {\bibinfo
  {journal} {Science bulletin}\ }\textbf {\bibinfo {volume} {66}},\ \bibinfo
  {pages} {1384} (\bibinfo {year} {2021})}\BibitemShut {NoStop}%
\bibitem [{\citenamefont {Yu}\ \emph {et~al.}(2021{\natexlab{a}})\citenamefont
  {Yu}, \citenamefont {Wang}, \citenamefont {Zhang}, \citenamefont {Sander},
  \citenamefont {Ni}, \citenamefont {Lu}, \citenamefont {Ma}, \citenamefont
  {Wang}, \citenamefont {Zhao}, \citenamefont {Chen} \emph
  {et~al.}}]{yu2021evidence}%
  \BibitemOpen
  \bibfield  {author} {\bibinfo {author} {\bibfnamefont {L.}~\bibnamefont
  {Yu}}, \bibinfo {author} {\bibfnamefont {C.}~\bibnamefont {Wang}}, \bibinfo
  {author} {\bibfnamefont {Y.}~\bibnamefont {Zhang}}, \bibinfo {author}
  {\bibfnamefont {M.}~\bibnamefont {Sander}}, \bibinfo {author} {\bibfnamefont
  {S.}~\bibnamefont {Ni}}, \bibinfo {author} {\bibfnamefont {Z.}~\bibnamefont
  {Lu}}, \bibinfo {author} {\bibfnamefont {S.}~\bibnamefont {Ma}}, \bibinfo
  {author} {\bibfnamefont {Z.}~\bibnamefont {Wang}}, \bibinfo {author}
  {\bibfnamefont {Z.}~\bibnamefont {Zhao}}, \bibinfo {author} {\bibfnamefont
  {H.}~\bibnamefont {Chen}},  \emph {et~al.},\ }\href@noop {} {\bibfield
  {journal} {\bibinfo  {journal} {arXiv preprint arXiv:2107.10714}\ } (\bibinfo
  {year} {2021}{\natexlab{a}})}\BibitemShut {NoStop}%
\bibitem [{\citenamefont {Yang}\ \emph {et~al.}(2020)\citenamefont {Yang},
  \citenamefont {Wang}, \citenamefont {Ortiz}, \citenamefont {Liu},
  \citenamefont {Gayles}, \citenamefont {Derunova}, \citenamefont
  {Gonzalez-Hernandez}, \citenamefont {{\v{S}}mejkal}, \citenamefont {Chen},
  \citenamefont {Parkin} \emph {et~al.}}]{yang2020giant}%
  \BibitemOpen
  \bibfield  {author} {\bibinfo {author} {\bibfnamefont {S.-Y.}\ \bibnamefont
  {Yang}}, \bibinfo {author} {\bibfnamefont {Y.}~\bibnamefont {Wang}}, \bibinfo
  {author} {\bibfnamefont {B.~R.}\ \bibnamefont {Ortiz}}, \bibinfo {author}
  {\bibfnamefont {D.}~\bibnamefont {Liu}}, \bibinfo {author} {\bibfnamefont
  {J.}~\bibnamefont {Gayles}}, \bibinfo {author} {\bibfnamefont
  {E.}~\bibnamefont {Derunova}}, \bibinfo {author} {\bibfnamefont
  {R.}~\bibnamefont {Gonzalez-Hernandez}}, \bibinfo {author} {\bibfnamefont
  {L.}~\bibnamefont {{\v{S}}mejkal}}, \bibinfo {author} {\bibfnamefont
  {Y.}~\bibnamefont {Chen}}, \bibinfo {author} {\bibfnamefont {S.~S.}\
  \bibnamefont {Parkin}},  \emph {et~al.},\ }\href
  {https://www.science.org/doi/full/10.1126/sciadv.abb6003} {\bibfield
  {journal} {\bibinfo  {journal} {Science advances}\ }\textbf {\bibinfo
  {volume} {6}},\ \bibinfo {pages} {eabb6003} (\bibinfo {year}
  {2020})}\BibitemShut {NoStop}%
\bibitem [{\citenamefont {Yu}\ \emph {et~al.}(2021{\natexlab{b}})\citenamefont
  {Yu}, \citenamefont {Wu}, \citenamefont {Wang}, \citenamefont {Lei},
  \citenamefont {Zhuo}, \citenamefont {Ying},\ and\ \citenamefont
  {Chen}}]{FHYuPrb2021}%
  \BibitemOpen
  \bibfield  {author} {\bibinfo {author} {\bibfnamefont {F.~H.}\ \bibnamefont
  {Yu}}, \bibinfo {author} {\bibfnamefont {T.}~\bibnamefont {Wu}}, \bibinfo
  {author} {\bibfnamefont {Z.~Y.}\ \bibnamefont {Wang}}, \bibinfo {author}
  {\bibfnamefont {B.}~\bibnamefont {Lei}}, \bibinfo {author} {\bibfnamefont
  {W.~Z.}\ \bibnamefont {Zhuo}}, \bibinfo {author} {\bibfnamefont {J.~J.}\
  \bibnamefont {Ying}}, \ and\ \bibinfo {author} {\bibfnamefont {X.~H.}\
  \bibnamefont {Chen}},\ }\href {\doibase 10.1103/PhysRevB.104.L041103}
  {\bibfield  {journal} {\bibinfo  {journal} {Phys. Rev. B}\ }\textbf {\bibinfo
  {volume} {104}},\ \bibinfo {pages} {L041103} (\bibinfo {year}
  {2021}{\natexlab{b}})}\BibitemShut {NoStop}%
\bibitem [{\citenamefont {Kenney}\ \emph {et~al.}(2021)\citenamefont {Kenney},
  \citenamefont {Ortiz}, \citenamefont {Wang}, \citenamefont {Wilson},\ and\
  \citenamefont {Graf}}]{kenney2021absence}%
  \BibitemOpen
  \bibfield  {author} {\bibinfo {author} {\bibfnamefont {E.~M.}\ \bibnamefont
  {Kenney}}, \bibinfo {author} {\bibfnamefont {B.~R.}\ \bibnamefont {Ortiz}},
  \bibinfo {author} {\bibfnamefont {C.}~\bibnamefont {Wang}}, \bibinfo {author}
  {\bibfnamefont {S.~D.}\ \bibnamefont {Wilson}}, \ and\ \bibinfo {author}
  {\bibfnamefont {M.~J.}\ \bibnamefont {Graf}},\ }\href
  {https://iopscience.iop.org/article/10.1088/1361-648X/abe8f9/meta} {\bibfield
   {journal} {\bibinfo  {journal} {Journal of Physics: Condensed Matter}\
  }\textbf {\bibinfo {volume} {33}},\ \bibinfo {pages} {235801} (\bibinfo
  {year} {2021})}\BibitemShut {NoStop}%
\bibitem [{\citenamefont {Mielke~III}\ \emph {et~al.}(2022)\citenamefont
  {Mielke~III}, \citenamefont {Das}, \citenamefont {Yin}, \citenamefont {Liu},
  \citenamefont {Gupta}, \citenamefont {Jiang}, \citenamefont {Medarde},
  \citenamefont {Wu}, \citenamefont {Lei}, \citenamefont {Chang} \emph
  {et~al.}}]{mielke2022time}%
  \BibitemOpen
  \bibfield  {author} {\bibinfo {author} {\bibfnamefont {C.}~\bibnamefont
  {Mielke~III}}, \bibinfo {author} {\bibfnamefont {D.}~\bibnamefont {Das}},
  \bibinfo {author} {\bibfnamefont {J.-X.}\ \bibnamefont {Yin}}, \bibinfo
  {author} {\bibfnamefont {H.}~\bibnamefont {Liu}}, \bibinfo {author}
  {\bibfnamefont {R.}~\bibnamefont {Gupta}}, \bibinfo {author} {\bibfnamefont
  {Y.-X.}\ \bibnamefont {Jiang}}, \bibinfo {author} {\bibfnamefont
  {M.}~\bibnamefont {Medarde}}, \bibinfo {author} {\bibfnamefont
  {X.}~\bibnamefont {Wu}}, \bibinfo {author} {\bibfnamefont {H.}~\bibnamefont
  {Lei}}, \bibinfo {author} {\bibfnamefont {J.}~\bibnamefont {Chang}},  \emph
  {et~al.},\ }\href {https://www.nature.com/articles/s41586-021-04327-z}
  {\bibfield  {journal} {\bibinfo  {journal} {Nature}\ }\textbf {\bibinfo
  {volume} {602}},\ \bibinfo {pages} {245} (\bibinfo {year}
  {2022})}\BibitemShut {NoStop}%
\bibitem [{\citenamefont {Khasanov}\ \emph {et~al.}(2022)\citenamefont
  {Khasanov}, \citenamefont {Das}, \citenamefont {Gupta}, \citenamefont
  {Mielke}, \citenamefont {Elender}, \citenamefont {Yin}, \citenamefont {Tu},
  \citenamefont {Gong}, \citenamefont {Lei}, \citenamefont {Ritz},
  \citenamefont {Fernandes}, \citenamefont {Birol}, \citenamefont {Guguchia},\
  and\ \citenamefont {Luetkens}}]{KhasanovPRRTR2022}%
  \BibitemOpen
  \bibfield  {author} {\bibinfo {author} {\bibfnamefont {R.}~\bibnamefont
  {Khasanov}}, \bibinfo {author} {\bibfnamefont {D.}~\bibnamefont {Das}},
  \bibinfo {author} {\bibfnamefont {R.}~\bibnamefont {Gupta}}, \bibinfo
  {author} {\bibfnamefont {C.}~\bibnamefont {Mielke}}, \bibinfo {author}
  {\bibfnamefont {M.}~\bibnamefont {Elender}}, \bibinfo {author} {\bibfnamefont
  {Q.}~\bibnamefont {Yin}}, \bibinfo {author} {\bibfnamefont {Z.}~\bibnamefont
  {Tu}}, \bibinfo {author} {\bibfnamefont {C.}~\bibnamefont {Gong}}, \bibinfo
  {author} {\bibfnamefont {H.}~\bibnamefont {Lei}}, \bibinfo {author}
  {\bibfnamefont {E.~T.}\ \bibnamefont {Ritz}}, \bibinfo {author}
  {\bibfnamefont {R.~M.}\ \bibnamefont {Fernandes}}, \bibinfo {author}
  {\bibfnamefont {T.}~\bibnamefont {Birol}}, \bibinfo {author} {\bibfnamefont
  {Z.}~\bibnamefont {Guguchia}}, \ and\ \bibinfo {author} {\bibfnamefont
  {H.}~\bibnamefont {Luetkens}},\ }\href {\doibase
  10.1103/PhysRevResearch.4.023244} {\bibfield  {journal} {\bibinfo  {journal}
  {Phys. Rev. Res.}\ }\textbf {\bibinfo {volume} {4}},\ \bibinfo {pages}
  {023244} (\bibinfo {year} {2022})}\BibitemShut {NoStop}%
\bibitem [{\citenamefont {Gupta}\ \emph {et~al.}(2022)\citenamefont {Gupta},
  \citenamefont {Das}, \citenamefont {Mielke~III}, \citenamefont {Ritz},
  \citenamefont {Hotz}, \citenamefont {Yin}, \citenamefont {Tu}, \citenamefont
  {Gong}, \citenamefont {Lei}, \citenamefont {Birol} \emph
  {et~al.}}]{gupta2022two}%
  \BibitemOpen
  \bibfield  {author} {\bibinfo {author} {\bibfnamefont {R.}~\bibnamefont
  {Gupta}}, \bibinfo {author} {\bibfnamefont {D.}~\bibnamefont {Das}}, \bibinfo
  {author} {\bibfnamefont {C.}~\bibnamefont {Mielke~III}}, \bibinfo {author}
  {\bibfnamefont {E.}~\bibnamefont {Ritz}}, \bibinfo {author} {\bibfnamefont
  {F.}~\bibnamefont {Hotz}}, \bibinfo {author} {\bibfnamefont {Q.}~\bibnamefont
  {Yin}}, \bibinfo {author} {\bibfnamefont {Z.}~\bibnamefont {Tu}}, \bibinfo
  {author} {\bibfnamefont {C.}~\bibnamefont {Gong}}, \bibinfo {author}
  {\bibfnamefont {H.}~\bibnamefont {Lei}}, \bibinfo {author} {\bibfnamefont
  {T.}~\bibnamefont {Birol}},  \emph {et~al.},\ }\href@noop {} {\bibfield
  {journal} {\bibinfo  {journal} {arXiv preprint arXiv:2203.05055}\ } (\bibinfo
  {year} {2022})}\BibitemShut {NoStop}%
\bibitem [{\citenamefont {Xu}\ \emph {et~al.}(2022)\citenamefont {Xu},
  \citenamefont {Ni}, \citenamefont {Liu}, \citenamefont {Ortiz}, \citenamefont
  {Deng}, \citenamefont {Wilson}, \citenamefont {Yan}, \citenamefont
  {Balents},\ and\ \citenamefont {Wu}}]{xu2022universal}%
  \BibitemOpen
  \bibfield  {author} {\bibinfo {author} {\bibfnamefont {Y.}~\bibnamefont
  {Xu}}, \bibinfo {author} {\bibfnamefont {Z.}~\bibnamefont {Ni}}, \bibinfo
  {author} {\bibfnamefont {Y.}~\bibnamefont {Liu}}, \bibinfo {author}
  {\bibfnamefont {B.~R.}\ \bibnamefont {Ortiz}}, \bibinfo {author}
  {\bibfnamefont {Q.}~\bibnamefont {Deng}}, \bibinfo {author} {\bibfnamefont
  {S.~D.}\ \bibnamefont {Wilson}}, \bibinfo {author} {\bibfnamefont
  {B.}~\bibnamefont {Yan}}, \bibinfo {author} {\bibfnamefont {L.}~\bibnamefont
  {Balents}}, \ and\ \bibinfo {author} {\bibfnamefont {L.}~\bibnamefont {Wu}},\
  }\href {\doibase 10.1038/s41567-022-01805-7} {\bibfield  {journal} {\bibinfo
  {journal} {Nature Physics}\ }\textbf {\bibinfo {volume} {18}},\ \bibinfo
  {pages} {1470} (\bibinfo {year} {2022})}\BibitemShut {NoStop}%
\bibitem [{\citenamefont {Lin}\ and\ \citenamefont
  {Nandkishore}(2021)}]{LinLoopprb2021}%
  \BibitemOpen
  \bibfield  {author} {\bibinfo {author} {\bibfnamefont {Y.-P.}\ \bibnamefont
  {Lin}}\ and\ \bibinfo {author} {\bibfnamefont {R.~M.}\ \bibnamefont
  {Nandkishore}},\ }\href {\doibase 10.1103/PhysRevB.104.045122} {\bibfield
  {journal} {\bibinfo  {journal} {Phys. Rev. B}\ }\textbf {\bibinfo {volume}
  {104}},\ \bibinfo {pages} {045122} (\bibinfo {year} {2021})}\BibitemShut
  {NoStop}%
\bibitem [{\citenamefont {Christensen}\ \emph {et~al.}(2022)\citenamefont
  {Christensen}, \citenamefont {Birol}, \citenamefont {Andersen},\ and\
  \citenamefont {Fernandes}}]{ChristensenloopPRB2022}%
  \BibitemOpen
  \bibfield  {author} {\bibinfo {author} {\bibfnamefont {M.~H.}\ \bibnamefont
  {Christensen}}, \bibinfo {author} {\bibfnamefont {T.}~\bibnamefont {Birol}},
  \bibinfo {author} {\bibfnamefont {B.~M.}\ \bibnamefont {Andersen}}, \ and\
  \bibinfo {author} {\bibfnamefont {R.~M.}\ \bibnamefont {Fernandes}},\ }\href
  {\doibase 10.1103/PhysRevB.106.144504} {\bibfield  {journal} {\bibinfo
  {journal} {Phys. Rev. B}\ }\textbf {\bibinfo {volume} {106}},\ \bibinfo
  {pages} {144504} (\bibinfo {year} {2022})}\BibitemShut {NoStop}%
\bibitem [{\citenamefont {Wang}\ \emph {et~al.}(2020)\citenamefont {Wang},
  \citenamefont {Yang}, \citenamefont {Sivakumar}, \citenamefont {Ortiz},
  \citenamefont {Teicher}, \citenamefont {Wu}, \citenamefont {Srivastava},
  \citenamefont {Garg}, \citenamefont {Liu}, \citenamefont {Parkin} \emph
  {et~al.}}]{wang2020proximity}%
  \BibitemOpen
  \bibfield  {author} {\bibinfo {author} {\bibfnamefont {Y.}~\bibnamefont
  {Wang}}, \bibinfo {author} {\bibfnamefont {S.}~\bibnamefont {Yang}}, \bibinfo
  {author} {\bibfnamefont {P.~K.}\ \bibnamefont {Sivakumar}}, \bibinfo {author}
  {\bibfnamefont {B.~R.}\ \bibnamefont {Ortiz}}, \bibinfo {author}
  {\bibfnamefont {S.~M.}\ \bibnamefont {Teicher}}, \bibinfo {author}
  {\bibfnamefont {H.}~\bibnamefont {Wu}}, \bibinfo {author} {\bibfnamefont
  {A.~K.}\ \bibnamefont {Srivastava}}, \bibinfo {author} {\bibfnamefont
  {C.}~\bibnamefont {Garg}}, \bibinfo {author} {\bibfnamefont {D.}~\bibnamefont
  {Liu}}, \bibinfo {author} {\bibfnamefont {S.~S.}\ \bibnamefont {Parkin}},
  \emph {et~al.},\ }\href@noop {} {\bibfield  {journal} {\bibinfo  {journal}
  {arXiv preprint arXiv:2012.05898}\ } (\bibinfo {year} {2020})}\BibitemShut
  {NoStop}%
\bibitem [{\citenamefont {Chen}\ \emph
  {et~al.}(2021{\natexlab{a}})\citenamefont {Chen}, \citenamefont {Wang},
  \citenamefont {Yin}, \citenamefont {Gu}, \citenamefont {Jiang}, \citenamefont
  {Tu}, \citenamefont {Gong}, \citenamefont {Uwatoko}, \citenamefont {Sun},
  \citenamefont {Lei}, \citenamefont {Hu},\ and\ \citenamefont
  {Cheng}}]{KChenPRL2021}%
  \BibitemOpen
  \bibfield  {author} {\bibinfo {author} {\bibfnamefont {K.~Y.}\ \bibnamefont
  {Chen}}, \bibinfo {author} {\bibfnamefont {N.~N.}\ \bibnamefont {Wang}},
  \bibinfo {author} {\bibfnamefont {Q.~W.}\ \bibnamefont {Yin}}, \bibinfo
  {author} {\bibfnamefont {Y.~H.}\ \bibnamefont {Gu}}, \bibinfo {author}
  {\bibfnamefont {K.}~\bibnamefont {Jiang}}, \bibinfo {author} {\bibfnamefont
  {Z.~J.}\ \bibnamefont {Tu}}, \bibinfo {author} {\bibfnamefont {C.~S.}\
  \bibnamefont {Gong}}, \bibinfo {author} {\bibfnamefont {Y.}~\bibnamefont
  {Uwatoko}}, \bibinfo {author} {\bibfnamefont {J.~P.}\ \bibnamefont {Sun}},
  \bibinfo {author} {\bibfnamefont {H.~C.}\ \bibnamefont {Lei}}, \bibinfo
  {author} {\bibfnamefont {J.~P.}\ \bibnamefont {Hu}}, \ and\ \bibinfo {author}
  {\bibfnamefont {J.-G.}\ \bibnamefont {Cheng}},\ }\href {\doibase
  10.1103/PhysRevLett.126.247001} {\bibfield  {journal} {\bibinfo  {journal}
  {Phys. Rev. Lett.}\ }\textbf {\bibinfo {volume} {126}},\ \bibinfo {pages}
  {247001} (\bibinfo {year} {2021}{\natexlab{a}})}\BibitemShut {NoStop}%
\bibitem [{\citenamefont {Ni}\ \emph {et~al.}(2021)\citenamefont {Ni},
  \citenamefont {Ma}, \citenamefont {Zhang}, \citenamefont {Yuan},
  \citenamefont {Yang}, \citenamefont {Lu}, \citenamefont {Wang}, \citenamefont
  {Sun}, \citenamefont {Zhao}, \citenamefont {Li} \emph
  {et~al.}}]{ni2021anisotropic}%
  \BibitemOpen
  \bibfield  {author} {\bibinfo {author} {\bibfnamefont {S.}~\bibnamefont
  {Ni}}, \bibinfo {author} {\bibfnamefont {S.}~\bibnamefont {Ma}}, \bibinfo
  {author} {\bibfnamefont {Y.}~\bibnamefont {Zhang}}, \bibinfo {author}
  {\bibfnamefont {J.}~\bibnamefont {Yuan}}, \bibinfo {author} {\bibfnamefont
  {H.}~\bibnamefont {Yang}}, \bibinfo {author} {\bibfnamefont {Z.}~\bibnamefont
  {Lu}}, \bibinfo {author} {\bibfnamefont {N.}~\bibnamefont {Wang}}, \bibinfo
  {author} {\bibfnamefont {J.}~\bibnamefont {Sun}}, \bibinfo {author}
  {\bibfnamefont {Z.}~\bibnamefont {Zhao}}, \bibinfo {author} {\bibfnamefont
  {D.}~\bibnamefont {Li}},  \emph {et~al.},\ }\href
  {https://iopscience.iop.org/article/10.1088/0256-307X/38/5/057403/meta}
  {\bibfield  {journal} {\bibinfo  {journal} {Chinese Physics Letters}\
  }\textbf {\bibinfo {volume} {38}},\ \bibinfo {pages} {057403} (\bibinfo
  {year} {2021})}\BibitemShut {NoStop}%
\bibitem [{\citenamefont {Mu}\ \emph {et~al.}(2021)\citenamefont {Mu},
  \citenamefont {Yin}, \citenamefont {Tu}, \citenamefont {Gong}, \citenamefont
  {Lei}, \citenamefont {Li},\ and\ \citenamefont {Luo}}]{mu2021s}%
  \BibitemOpen
  \bibfield  {author} {\bibinfo {author} {\bibfnamefont {C.}~\bibnamefont
  {Mu}}, \bibinfo {author} {\bibfnamefont {Q.}~\bibnamefont {Yin}}, \bibinfo
  {author} {\bibfnamefont {Z.}~\bibnamefont {Tu}}, \bibinfo {author}
  {\bibfnamefont {C.}~\bibnamefont {Gong}}, \bibinfo {author} {\bibfnamefont
  {H.}~\bibnamefont {Lei}}, \bibinfo {author} {\bibfnamefont {Z.}~\bibnamefont
  {Li}}, \ and\ \bibinfo {author} {\bibfnamefont {J.}~\bibnamefont {Luo}},\
  }\href
  {https://iopscience.iop.org/article/10.1088/0256-307X/38/7/077402/meta}
  {\bibfield  {journal} {\bibinfo  {journal} {Chinese Physics Letters}\
  }\textbf {\bibinfo {volume} {38}},\ \bibinfo {pages} {077402} (\bibinfo
  {year} {2021})}\BibitemShut {NoStop}%
\bibitem [{\citenamefont {Duan}\ \emph {et~al.}(2021)\citenamefont {Duan},
  \citenamefont {Nie}, \citenamefont {Luo}, \citenamefont {Yu}, \citenamefont
  {Ortiz}, \citenamefont {Yin}, \citenamefont {Su}, \citenamefont {Du},
  \citenamefont {Wang}, \citenamefont {Chen} \emph
  {et~al.}}]{duan2021nodeless}%
  \BibitemOpen
  \bibfield  {author} {\bibinfo {author} {\bibfnamefont {W.}~\bibnamefont
  {Duan}}, \bibinfo {author} {\bibfnamefont {Z.}~\bibnamefont {Nie}}, \bibinfo
  {author} {\bibfnamefont {S.}~\bibnamefont {Luo}}, \bibinfo {author}
  {\bibfnamefont {F.}~\bibnamefont {Yu}}, \bibinfo {author} {\bibfnamefont
  {B.~R.}\ \bibnamefont {Ortiz}}, \bibinfo {author} {\bibfnamefont
  {L.}~\bibnamefont {Yin}}, \bibinfo {author} {\bibfnamefont {H.}~\bibnamefont
  {Su}}, \bibinfo {author} {\bibfnamefont {F.}~\bibnamefont {Du}}, \bibinfo
  {author} {\bibfnamefont {A.}~\bibnamefont {Wang}}, \bibinfo {author}
  {\bibfnamefont {Y.}~\bibnamefont {Chen}},  \emph {et~al.},\ }\href
  {https://link.springer.com/article/10.1007/s11433-021-1747-7} {\bibfield
  {journal} {\bibinfo  {journal} {Science China Physics, Mechanics \&
  Astronomy}\ }\textbf {\bibinfo {volume} {64}},\ \bibinfo {pages} {107462}
  (\bibinfo {year} {2021})}\BibitemShut {NoStop}%
\bibitem [{\citenamefont {Zhao}\ \emph
  {et~al.}(2021{\natexlab{a}})\citenamefont {Zhao}, \citenamefont {Wang},
  \citenamefont {Xia}, \citenamefont {Yin}, \citenamefont {Ni}, \citenamefont
  {Huang}, \citenamefont {Tu}, \citenamefont {Tao}, \citenamefont {Tu},
  \citenamefont {Gong} \emph {et~al.}}]{zhao2021nodal}%
  \BibitemOpen
  \bibfield  {author} {\bibinfo {author} {\bibfnamefont {C.}~\bibnamefont
  {Zhao}}, \bibinfo {author} {\bibfnamefont {L.}~\bibnamefont {Wang}}, \bibinfo
  {author} {\bibfnamefont {W.}~\bibnamefont {Xia}}, \bibinfo {author}
  {\bibfnamefont {Q.}~\bibnamefont {Yin}}, \bibinfo {author} {\bibfnamefont
  {J.}~\bibnamefont {Ni}}, \bibinfo {author} {\bibfnamefont {Y.}~\bibnamefont
  {Huang}}, \bibinfo {author} {\bibfnamefont {C.}~\bibnamefont {Tu}}, \bibinfo
  {author} {\bibfnamefont {Z.}~\bibnamefont {Tao}}, \bibinfo {author}
  {\bibfnamefont {Z.}~\bibnamefont {Tu}}, \bibinfo {author} {\bibfnamefont
  {C.}~\bibnamefont {Gong}},  \emph {et~al.},\ }\href@noop {} {\bibfield
  {journal} {\bibinfo  {journal} {arXiv preprint arXiv:2102.08356}\ } (\bibinfo
  {year} {2021}{\natexlab{a}})}\BibitemShut {NoStop}%
\bibitem [{\citenamefont {Kiesel}\ \emph {et~al.}(2013)\citenamefont {Kiesel},
  \citenamefont {Platt},\ and\ \citenamefont {Thomale}}]{MKieselPRL2013}%
  \BibitemOpen
  \bibfield  {author} {\bibinfo {author} {\bibfnamefont {M.~L.}\ \bibnamefont
  {Kiesel}}, \bibinfo {author} {\bibfnamefont {C.}~\bibnamefont {Platt}}, \
  and\ \bibinfo {author} {\bibfnamefont {R.}~\bibnamefont {Thomale}},\ }\href
  {\doibase 10.1103/PhysRevLett.110.126405} {\bibfield  {journal} {\bibinfo
  {journal} {Phys. Rev. Lett.}\ }\textbf {\bibinfo {volume} {110}},\ \bibinfo
  {pages} {126405} (\bibinfo {year} {2013})}\BibitemShut {NoStop}%
\bibitem [{\citenamefont {Wang}\ \emph
  {et~al.}(2013{\natexlab{a}})\citenamefont {Wang}, \citenamefont {Li},
  \citenamefont {Xiang},\ and\ \citenamefont {Wang}}]{WangPRB2013}%
  \BibitemOpen
  \bibfield  {author} {\bibinfo {author} {\bibfnamefont {W.-S.}\ \bibnamefont
  {Wang}}, \bibinfo {author} {\bibfnamefont {Z.-Z.}\ \bibnamefont {Li}},
  \bibinfo {author} {\bibfnamefont {Y.-Y.}\ \bibnamefont {Xiang}}, \ and\
  \bibinfo {author} {\bibfnamefont {Q.-H.}\ \bibnamefont {Wang}},\ }\href
  {\doibase 10.1103/PhysRevB.87.115135} {\bibfield  {journal} {\bibinfo
  {journal} {Phys. Rev. B}\ }\textbf {\bibinfo {volume} {87}},\ \bibinfo
  {pages} {115135} (\bibinfo {year} {2013}{\natexlab{a}})}\BibitemShut
  {NoStop}%
\bibitem [{\citenamefont {Wu}\ \emph {et~al.}(2021)\citenamefont {Wu},
  \citenamefont {Schwemmer}, \citenamefont {M\"uller}, \citenamefont
  {Consiglio}, \citenamefont {Sangiovanni}, \citenamefont {Di~Sante},
  \citenamefont {Iqbal}, \citenamefont {Hanke}, \citenamefont {Schnyder},
  \citenamefont {Denner}, \citenamefont {Fischer}, \citenamefont {Neupert},\
  and\ \citenamefont {Thomale}}]{WuPRL2021}%
  \BibitemOpen
  \bibfield  {author} {\bibinfo {author} {\bibfnamefont {X.}~\bibnamefont
  {Wu}}, \bibinfo {author} {\bibfnamefont {T.}~\bibnamefont {Schwemmer}},
  \bibinfo {author} {\bibfnamefont {T.}~\bibnamefont {M\"uller}}, \bibinfo
  {author} {\bibfnamefont {A.}~\bibnamefont {Consiglio}}, \bibinfo {author}
  {\bibfnamefont {G.}~\bibnamefont {Sangiovanni}}, \bibinfo {author}
  {\bibfnamefont {D.}~\bibnamefont {Di~Sante}}, \bibinfo {author}
  {\bibfnamefont {Y.}~\bibnamefont {Iqbal}}, \bibinfo {author} {\bibfnamefont
  {W.}~\bibnamefont {Hanke}}, \bibinfo {author} {\bibfnamefont {A.~P.}\
  \bibnamefont {Schnyder}}, \bibinfo {author} {\bibfnamefont {M.~M.}\
  \bibnamefont {Denner}}, \bibinfo {author} {\bibfnamefont {M.~H.}\
  \bibnamefont {Fischer}}, \bibinfo {author} {\bibfnamefont {T.}~\bibnamefont
  {Neupert}}, \ and\ \bibinfo {author} {\bibfnamefont {R.}~\bibnamefont
  {Thomale}},\ }\href {\doibase 10.1103/PhysRevLett.127.177001} {\bibfield
  {journal} {\bibinfo  {journal} {Phys. Rev. Lett.}\ }\textbf {\bibinfo
  {volume} {127}},\ \bibinfo {pages} {177001} (\bibinfo {year}
  {2021})}\BibitemShut {NoStop}%
\bibitem [{\citenamefont {Wen}\ \emph {et~al.}(2022)\citenamefont {Wen},
  \citenamefont {Zhu}, \citenamefont {Xiao}, \citenamefont {Hao}, \citenamefont
  {Mondaini}, \citenamefont {Guo},\ and\ \citenamefont {Feng}}]{WenPRB2022}%
  \BibitemOpen
  \bibfield  {author} {\bibinfo {author} {\bibfnamefont {C.}~\bibnamefont
  {Wen}}, \bibinfo {author} {\bibfnamefont {X.}~\bibnamefont {Zhu}}, \bibinfo
  {author} {\bibfnamefont {Z.}~\bibnamefont {Xiao}}, \bibinfo {author}
  {\bibfnamefont {N.}~\bibnamefont {Hao}}, \bibinfo {author} {\bibfnamefont
  {R.}~\bibnamefont {Mondaini}}, \bibinfo {author} {\bibfnamefont
  {H.}~\bibnamefont {Guo}}, \ and\ \bibinfo {author} {\bibfnamefont
  {S.}~\bibnamefont {Feng}},\ }\href {\doibase 10.1103/PhysRevB.105.075118}
  {\bibfield  {journal} {\bibinfo  {journal} {Phys. Rev. B}\ }\textbf {\bibinfo
  {volume} {105}},\ \bibinfo {pages} {075118} (\bibinfo {year}
  {2022})}\BibitemShut {NoStop}%
\bibitem [{\citenamefont {Lin}\ and\ \citenamefont
  {Nandkishore}(2022{\natexlab{a}})}]{lin2022multidome}%
  \BibitemOpen
  \bibfield  {author} {\bibinfo {author} {\bibfnamefont {Y.-P.}\ \bibnamefont
  {Lin}}\ and\ \bibinfo {author} {\bibfnamefont {R.~M.}\ \bibnamefont
  {Nandkishore}},\ }\href
  {https://journals.aps.org/prb/abstract/10.1103/PhysRevB.106.L060507}
  {\bibfield  {journal} {\bibinfo  {journal} {Physical Review B}\ }\textbf
  {\bibinfo {volume} {106}},\ \bibinfo {pages} {L060507} (\bibinfo {year}
  {2022}{\natexlab{a}})}\BibitemShut {NoStop}%
\bibitem [{\citenamefont {Chen}\ \emph
  {et~al.}(2021{\natexlab{b}})\citenamefont {Chen}, \citenamefont {Yang},
  \citenamefont {Hu}, \citenamefont {Zhao}, \citenamefont {Yuan}, \citenamefont
  {Xing}, \citenamefont {Qian}, \citenamefont {Huang}, \citenamefont {Li},
  \citenamefont {Ye} \emph {et~al.}}]{chen2021roton}%
  \BibitemOpen
  \bibfield  {author} {\bibinfo {author} {\bibfnamefont {H.}~\bibnamefont
  {Chen}}, \bibinfo {author} {\bibfnamefont {H.}~\bibnamefont {Yang}}, \bibinfo
  {author} {\bibfnamefont {B.}~\bibnamefont {Hu}}, \bibinfo {author}
  {\bibfnamefont {Z.}~\bibnamefont {Zhao}}, \bibinfo {author} {\bibfnamefont
  {J.}~\bibnamefont {Yuan}}, \bibinfo {author} {\bibfnamefont {Y.}~\bibnamefont
  {Xing}}, \bibinfo {author} {\bibfnamefont {G.}~\bibnamefont {Qian}}, \bibinfo
  {author} {\bibfnamefont {Z.}~\bibnamefont {Huang}}, \bibinfo {author}
  {\bibfnamefont {G.}~\bibnamefont {Li}}, \bibinfo {author} {\bibfnamefont
  {Y.}~\bibnamefont {Ye}},  \emph {et~al.},\ }\href
  {https://www.nature.com/articles/s41586-021-03983-5} {\bibfield  {journal}
  {\bibinfo  {journal} {Nature}\ }\textbf {\bibinfo {volume} {599}},\ \bibinfo
  {pages} {222} (\bibinfo {year} {2021}{\natexlab{b}})}\BibitemShut {NoStop}%
\bibitem [{\citenamefont {Zhou}\ and\ \citenamefont
  {Wang}(2022)}]{zhou2022chern}%
  \BibitemOpen
  \bibfield  {author} {\bibinfo {author} {\bibfnamefont {S.}~\bibnamefont
  {Zhou}}\ and\ \bibinfo {author} {\bibfnamefont {Z.}~\bibnamefont {Wang}},\
  }\href {https://www.nature.com/articles/s41467-022-34832-2} {\bibfield
  {journal} {\bibinfo  {journal} {Nature Communications}\ }\textbf {\bibinfo
  {volume} {13}},\ \bibinfo {pages} {7288} (\bibinfo {year}
  {2022})}\BibitemShut {NoStop}%
\bibitem [{\citenamefont {Xiang}\ \emph {et~al.}(2021)\citenamefont {Xiang},
  \citenamefont {Li}, \citenamefont {Li}, \citenamefont {Xie}, \citenamefont
  {Yang}, \citenamefont {Wang}, \citenamefont {Yao},\ and\ \citenamefont
  {Wen}}]{xiang2021twofold}%
  \BibitemOpen
  \bibfield  {author} {\bibinfo {author} {\bibfnamefont {Y.}~\bibnamefont
  {Xiang}}, \bibinfo {author} {\bibfnamefont {Q.}~\bibnamefont {Li}}, \bibinfo
  {author} {\bibfnamefont {Y.}~\bibnamefont {Li}}, \bibinfo {author}
  {\bibfnamefont {W.}~\bibnamefont {Xie}}, \bibinfo {author} {\bibfnamefont
  {H.}~\bibnamefont {Yang}}, \bibinfo {author} {\bibfnamefont {Z.}~\bibnamefont
  {Wang}}, \bibinfo {author} {\bibfnamefont {Y.}~\bibnamefont {Yao}}, \ and\
  \bibinfo {author} {\bibfnamefont {H.-H.}\ \bibnamefont {Wen}},\ }\href
  {https://www.nature.com/articles/s41467-021-27084-z} {\bibfield  {journal}
  {\bibinfo  {journal} {Nature communications}\ }\textbf {\bibinfo {volume}
  {12}},\ \bibinfo {pages} {6727} (\bibinfo {year} {2021})}\BibitemShut
  {NoStop}%
\bibitem [{\citenamefont {Nie}\ \emph {et~al.}(2022)\citenamefont {Nie},
  \citenamefont {Sun}, \citenamefont {Ma}, \citenamefont {Song}, \citenamefont
  {Zheng}, \citenamefont {Liang}, \citenamefont {Wu}, \citenamefont {Yu},
  \citenamefont {Li}, \citenamefont {Shan} \emph {et~al.}}]{nie2022charge}%
  \BibitemOpen
  \bibfield  {author} {\bibinfo {author} {\bibfnamefont {L.}~\bibnamefont
  {Nie}}, \bibinfo {author} {\bibfnamefont {K.}~\bibnamefont {Sun}}, \bibinfo
  {author} {\bibfnamefont {W.}~\bibnamefont {Ma}}, \bibinfo {author}
  {\bibfnamefont {D.}~\bibnamefont {Song}}, \bibinfo {author} {\bibfnamefont
  {L.}~\bibnamefont {Zheng}}, \bibinfo {author} {\bibfnamefont
  {Z.}~\bibnamefont {Liang}}, \bibinfo {author} {\bibfnamefont
  {P.}~\bibnamefont {Wu}}, \bibinfo {author} {\bibfnamefont {F.}~\bibnamefont
  {Yu}}, \bibinfo {author} {\bibfnamefont {J.}~\bibnamefont {Li}}, \bibinfo
  {author} {\bibfnamefont {M.}~\bibnamefont {Shan}},  \emph {et~al.},\ }\href
  {https://www.nature.com/articles/s41586-022-04493-8} {\bibfield  {journal}
  {\bibinfo  {journal} {Nature}\ }\textbf {\bibinfo {volume} {604}},\ \bibinfo
  {pages} {59} (\bibinfo {year} {2022})}\BibitemShut {NoStop}%
\bibitem [{\citenamefont {Grandi}\ \emph {et~al.}(2023)\citenamefont {Grandi},
  \citenamefont {Consiglio}, \citenamefont {Sentef}, \citenamefont {Thomale},\
  and\ \citenamefont {Kennes}}]{grandi2023theory}%
  \BibitemOpen
  \bibfield  {author} {\bibinfo {author} {\bibfnamefont {F.}~\bibnamefont
  {Grandi}}, \bibinfo {author} {\bibfnamefont {A.}~\bibnamefont {Consiglio}},
  \bibinfo {author} {\bibfnamefont {M.~A.}\ \bibnamefont {Sentef}}, \bibinfo
  {author} {\bibfnamefont {R.}~\bibnamefont {Thomale}}, \ and\ \bibinfo
  {author} {\bibfnamefont {D.~M.}\ \bibnamefont {Kennes}},\ }\href {\doibase
  10.1103/PhysRevB.107.155131} {\bibfield  {journal} {\bibinfo  {journal}
  {Phys. Rev. B}\ }\textbf {\bibinfo {volume} {107}},\ \bibinfo {pages}
  {155131} (\bibinfo {year} {2023})}\BibitemShut {NoStop}%
\bibitem [{\citenamefont {Kang}\ \emph {et~al.}(2022)\citenamefont {Kang},
  \citenamefont {Fang}, \citenamefont {Kim}, \citenamefont {Ortiz},
  \citenamefont {Ryu}, \citenamefont {Kim}, \citenamefont {Yoo}, \citenamefont
  {Sangiovanni}, \citenamefont {Di~Sante}, \citenamefont {Park} \emph
  {et~al.}}]{kang2022twofold}%
  \BibitemOpen
  \bibfield  {author} {\bibinfo {author} {\bibfnamefont {M.}~\bibnamefont
  {Kang}}, \bibinfo {author} {\bibfnamefont {S.}~\bibnamefont {Fang}}, \bibinfo
  {author} {\bibfnamefont {J.-K.}\ \bibnamefont {Kim}}, \bibinfo {author}
  {\bibfnamefont {B.~R.}\ \bibnamefont {Ortiz}}, \bibinfo {author}
  {\bibfnamefont {S.~H.}\ \bibnamefont {Ryu}}, \bibinfo {author} {\bibfnamefont
  {J.}~\bibnamefont {Kim}}, \bibinfo {author} {\bibfnamefont {J.}~\bibnamefont
  {Yoo}}, \bibinfo {author} {\bibfnamefont {G.}~\bibnamefont {Sangiovanni}},
  \bibinfo {author} {\bibfnamefont {D.}~\bibnamefont {Di~Sante}}, \bibinfo
  {author} {\bibfnamefont {B.-G.}\ \bibnamefont {Park}},  \emph {et~al.},\
  }\href {https://www.nature.com/articles/s41567-021-01451-5} {\bibfield
  {journal} {\bibinfo  {journal} {Nature Physics}\ }\textbf {\bibinfo {volume}
  {18}},\ \bibinfo {pages} {301} (\bibinfo {year} {2022})}\BibitemShut
  {NoStop}%
\bibitem [{\citenamefont {Lou}\ \emph {et~al.}(2022)\citenamefont {Lou},
  \citenamefont {Fedorov}, \citenamefont {Yin}, \citenamefont {Kuibarov},
  \citenamefont {Tu}, \citenamefont {Gong}, \citenamefont {Schwier},
  \citenamefont {B\"uchner}, \citenamefont {Lei},\ and\ \citenamefont
  {Borisenko}}]{Louupeakdiphumpprl2022}%
  \BibitemOpen
  \bibfield  {author} {\bibinfo {author} {\bibfnamefont {R.}~\bibnamefont
  {Lou}}, \bibinfo {author} {\bibfnamefont {A.}~\bibnamefont {Fedorov}},
  \bibinfo {author} {\bibfnamefont {Q.}~\bibnamefont {Yin}}, \bibinfo {author}
  {\bibfnamefont {A.}~\bibnamefont {Kuibarov}}, \bibinfo {author}
  {\bibfnamefont {Z.}~\bibnamefont {Tu}}, \bibinfo {author} {\bibfnamefont
  {C.}~\bibnamefont {Gong}}, \bibinfo {author} {\bibfnamefont {E.~F.}\
  \bibnamefont {Schwier}}, \bibinfo {author} {\bibfnamefont {B.}~\bibnamefont
  {B\"uchner}}, \bibinfo {author} {\bibfnamefont {H.}~\bibnamefont {Lei}}, \
  and\ \bibinfo {author} {\bibfnamefont {S.}~\bibnamefont {Borisenko}},\ }\href
  {\doibase 10.1103/PhysRevLett.128.036402} {\bibfield  {journal} {\bibinfo
  {journal} {Phys. Rev. Lett.}\ }\textbf {\bibinfo {volume} {128}},\ \bibinfo
  {pages} {036402} (\bibinfo {year} {2022})}\BibitemShut {NoStop}%
\bibitem [{\citenamefont {Luo}\ \emph {et~al.}(2022{\natexlab{a}})\citenamefont
  {Luo}, \citenamefont {Gao}, \citenamefont {Liu}, \citenamefont {Gu},
  \citenamefont {Wu}, \citenamefont {Yi}, \citenamefont {Jia}, \citenamefont
  {Wu}, \citenamefont {Luo}, \citenamefont {Xu} \emph
  {et~al.}}]{luo2022electronic}%
  \BibitemOpen
  \bibfield  {author} {\bibinfo {author} {\bibfnamefont {H.}~\bibnamefont
  {Luo}}, \bibinfo {author} {\bibfnamefont {Q.}~\bibnamefont {Gao}}, \bibinfo
  {author} {\bibfnamefont {H.}~\bibnamefont {Liu}}, \bibinfo {author}
  {\bibfnamefont {Y.}~\bibnamefont {Gu}}, \bibinfo {author} {\bibfnamefont
  {D.}~\bibnamefont {Wu}}, \bibinfo {author} {\bibfnamefont {C.}~\bibnamefont
  {Yi}}, \bibinfo {author} {\bibfnamefont {J.}~\bibnamefont {Jia}}, \bibinfo
  {author} {\bibfnamefont {S.}~\bibnamefont {Wu}}, \bibinfo {author}
  {\bibfnamefont {X.}~\bibnamefont {Luo}}, \bibinfo {author} {\bibfnamefont
  {Y.}~\bibnamefont {Xu}},  \emph {et~al.},\ }\href
  {https://www.nature.com/articles/s41467-021-27946-6} {\bibfield  {journal}
  {\bibinfo  {journal} {Nature communications}\ }\textbf {\bibinfo {volume}
  {13}},\ \bibinfo {pages} {273} (\bibinfo {year}
  {2022}{\natexlab{a}})}\BibitemShut {NoStop}%
\bibitem [{\citenamefont {Wu}\ \emph {et~al.}(2022)\citenamefont {Wu},
  \citenamefont {Ortiz}, \citenamefont {Tan}, \citenamefont {Wilson},
  \citenamefont {Yan}, \citenamefont {Birol},\ and\ \citenamefont
  {Blumberg}}]{Wuprb2022}%
  \BibitemOpen
  \bibfield  {author} {\bibinfo {author} {\bibfnamefont {S.}~\bibnamefont
  {Wu}}, \bibinfo {author} {\bibfnamefont {B.~R.}\ \bibnamefont {Ortiz}},
  \bibinfo {author} {\bibfnamefont {H.}~\bibnamefont {Tan}}, \bibinfo {author}
  {\bibfnamefont {S.~D.}\ \bibnamefont {Wilson}}, \bibinfo {author}
  {\bibfnamefont {B.}~\bibnamefont {Yan}}, \bibinfo {author} {\bibfnamefont
  {T.}~\bibnamefont {Birol}}, \ and\ \bibinfo {author} {\bibfnamefont
  {G.}~\bibnamefont {Blumberg}},\ }\href {\doibase 10.1103/PhysRevB.105.155106}
  {\bibfield  {journal} {\bibinfo  {journal} {Phys. Rev. B}\ }\textbf {\bibinfo
  {volume} {105}},\ \bibinfo {pages} {155106} (\bibinfo {year}
  {2022})}\BibitemShut {NoStop}%
\bibitem [{\citenamefont {Tazai}\ \emph {et~al.}(2022)\citenamefont {Tazai},
  \citenamefont {Yamakawa}, \citenamefont {Onari},\ and\ \citenamefont
  {Kontani}}]{tazai2022mechanism}%
  \BibitemOpen
  \bibfield  {author} {\bibinfo {author} {\bibfnamefont {R.}~\bibnamefont
  {Tazai}}, \bibinfo {author} {\bibfnamefont {Y.}~\bibnamefont {Yamakawa}},
  \bibinfo {author} {\bibfnamefont {S.}~\bibnamefont {Onari}}, \ and\ \bibinfo
  {author} {\bibfnamefont {H.}~\bibnamefont {Kontani}},\ }\href
  {https://www.science.org/doi/abs/10.1126/sciadv.abl4108} {\bibfield
  {journal} {\bibinfo  {journal} {Science Advances}\ }\textbf {\bibinfo
  {volume} {8}},\ \bibinfo {pages} {eabl4108} (\bibinfo {year}
  {2022})}\BibitemShut {NoStop}%
\bibitem [{\citenamefont {Feng}\ \emph {et~al.}(2023)\citenamefont {Feng},
  \citenamefont {Zhao}, \citenamefont {Luo}, \citenamefont {Yang},
  \citenamefont {Fang}, \citenamefont {Yang}, \citenamefont {Gao},
  \citenamefont {Zhou},\ and\ \citenamefont {Zheng}}]{feng2023commensurate}%
  \BibitemOpen
  \bibfield  {author} {\bibinfo {author} {\bibfnamefont {X.}~\bibnamefont
  {Feng}}, \bibinfo {author} {\bibfnamefont {Z.}~\bibnamefont {Zhao}}, \bibinfo
  {author} {\bibfnamefont {J.}~\bibnamefont {Luo}}, \bibinfo {author}
  {\bibfnamefont {J.}~\bibnamefont {Yang}}, \bibinfo {author} {\bibfnamefont
  {A.}~\bibnamefont {Fang}}, \bibinfo {author} {\bibfnamefont {H.}~\bibnamefont
  {Yang}}, \bibinfo {author} {\bibfnamefont {H.}~\bibnamefont {Gao}}, \bibinfo
  {author} {\bibfnamefont {R.}~\bibnamefont {Zhou}}, \ and\ \bibinfo {author}
  {\bibfnamefont {G.-q.}\ \bibnamefont {Zheng}},\ }\href@noop {} {\bibfield
  {journal} {\bibinfo  {journal} {arXiv preprint arXiv:2303.01225}\ } (\bibinfo
  {year} {2023})}\BibitemShut {NoStop}%
\bibitem [{\citenamefont {Luo}\ \emph {et~al.}(2022{\natexlab{b}})\citenamefont
  {Luo}, \citenamefont {Zhao}, \citenamefont {Zhou}, \citenamefont {Yang},
  \citenamefont {Fang}, \citenamefont {Yang}, \citenamefont {Gao},
  \citenamefont {Zhou},\ and\ \citenamefont {Zheng}}]{luo2022possible}%
  \BibitemOpen
  \bibfield  {author} {\bibinfo {author} {\bibfnamefont {J.}~\bibnamefont
  {Luo}}, \bibinfo {author} {\bibfnamefont {Z.}~\bibnamefont {Zhao}}, \bibinfo
  {author} {\bibfnamefont {Y.}~\bibnamefont {Zhou}}, \bibinfo {author}
  {\bibfnamefont {J.}~\bibnamefont {Yang}}, \bibinfo {author} {\bibfnamefont
  {A.}~\bibnamefont {Fang}}, \bibinfo {author} {\bibfnamefont {H.}~\bibnamefont
  {Yang}}, \bibinfo {author} {\bibfnamefont {H.}~\bibnamefont {Gao}}, \bibinfo
  {author} {\bibfnamefont {R.}~\bibnamefont {Zhou}}, \ and\ \bibinfo {author}
  {\bibfnamefont {G.-q.}\ \bibnamefont {Zheng}},\ }\href
  {https://www.nature.com/articles/s41535-022-00437-7} {\bibfield  {journal}
  {\bibinfo  {journal} {npj Quantum Materials}\ }\textbf {\bibinfo {volume}
  {7}},\ \bibinfo {pages} {30} (\bibinfo {year}
  {2022}{\natexlab{b}})}\BibitemShut {NoStop}%
\bibitem [{\citenamefont {Hu}\ \emph {et~al.}(2022{\natexlab{a}})\citenamefont
  {Hu}, \citenamefont {Wu}, \citenamefont {Ortiz}, \citenamefont {Han},
  \citenamefont {Plumb}, \citenamefont {Wilson}, \citenamefont {Schnyder},\
  and\ \citenamefont {Shi}}]{HuPRB2022}%
  \BibitemOpen
  \bibfield  {author} {\bibinfo {author} {\bibfnamefont {Y.}~\bibnamefont
  {Hu}}, \bibinfo {author} {\bibfnamefont {X.}~\bibnamefont {Wu}}, \bibinfo
  {author} {\bibfnamefont {B.~R.}\ \bibnamefont {Ortiz}}, \bibinfo {author}
  {\bibfnamefont {X.}~\bibnamefont {Han}}, \bibinfo {author} {\bibfnamefont
  {N.~C.}\ \bibnamefont {Plumb}}, \bibinfo {author} {\bibfnamefont {S.~D.}\
  \bibnamefont {Wilson}}, \bibinfo {author} {\bibfnamefont {A.~P.}\
  \bibnamefont {Schnyder}}, \ and\ \bibinfo {author} {\bibfnamefont
  {M.}~\bibnamefont {Shi}},\ }\href {\doibase 10.1103/PhysRevB.106.L241106}
  {\bibfield  {journal} {\bibinfo  {journal} {Phys. Rev. B}\ }\textbf {\bibinfo
  {volume} {106}},\ \bibinfo {pages} {L241106} (\bibinfo {year}
  {2022}{\natexlab{a}})}\BibitemShut {NoStop}%
\bibitem [{\citenamefont {Cho}\ \emph {et~al.}(2021)\citenamefont {Cho},
  \citenamefont {Ma}, \citenamefont {Xia}, \citenamefont {Yang}, \citenamefont
  {Liu}, \citenamefont {Huang}, \citenamefont {Jiang}, \citenamefont {Lu},
  \citenamefont {Liu}, \citenamefont {Liu}, \citenamefont {Li}, \citenamefont
  {Wang}, \citenamefont {Liu}, \citenamefont {Jia}, \citenamefont {Guo},
  \citenamefont {Liu},\ and\ \citenamefont {Shen}}]{ChoARPESPRL2021}%
  \BibitemOpen
  \bibfield  {author} {\bibinfo {author} {\bibfnamefont {S.}~\bibnamefont
  {Cho}}, \bibinfo {author} {\bibfnamefont {H.}~\bibnamefont {Ma}}, \bibinfo
  {author} {\bibfnamefont {W.}~\bibnamefont {Xia}}, \bibinfo {author}
  {\bibfnamefont {Y.}~\bibnamefont {Yang}}, \bibinfo {author} {\bibfnamefont
  {Z.}~\bibnamefont {Liu}}, \bibinfo {author} {\bibfnamefont {Z.}~\bibnamefont
  {Huang}}, \bibinfo {author} {\bibfnamefont {Z.}~\bibnamefont {Jiang}},
  \bibinfo {author} {\bibfnamefont {X.}~\bibnamefont {Lu}}, \bibinfo {author}
  {\bibfnamefont {J.}~\bibnamefont {Liu}}, \bibinfo {author} {\bibfnamefont
  {Z.}~\bibnamefont {Liu}}, \bibinfo {author} {\bibfnamefont {J.}~\bibnamefont
  {Li}}, \bibinfo {author} {\bibfnamefont {J.}~\bibnamefont {Wang}}, \bibinfo
  {author} {\bibfnamefont {Y.}~\bibnamefont {Liu}}, \bibinfo {author}
  {\bibfnamefont {J.}~\bibnamefont {Jia}}, \bibinfo {author} {\bibfnamefont
  {Y.}~\bibnamefont {Guo}}, \bibinfo {author} {\bibfnamefont {J.}~\bibnamefont
  {Liu}}, \ and\ \bibinfo {author} {\bibfnamefont {D.}~\bibnamefont {Shen}},\
  }\href {\doibase 10.1103/PhysRevLett.127.236401} {\bibfield  {journal}
  {\bibinfo  {journal} {Phys. Rev. Lett.}\ }\textbf {\bibinfo {volume} {127}},\
  \bibinfo {pages} {236401} (\bibinfo {year} {2021})}\BibitemShut {NoStop}%
\bibitem [{\citenamefont {Li}\ \emph {et~al.}(2022)\citenamefont {Li},
  \citenamefont {Fabbris}, \citenamefont {Said}, \citenamefont {Pai},
  \citenamefont {Yin}, \citenamefont {Gong}, \citenamefont {Tu}, \citenamefont
  {Lei}, \citenamefont {Sun}, \citenamefont {Cheng} \emph
  {et~al.}}]{li2022conjoined}%
  \BibitemOpen
  \bibfield  {author} {\bibinfo {author} {\bibfnamefont {H.}~\bibnamefont
  {Li}}, \bibinfo {author} {\bibfnamefont {G.}~\bibnamefont {Fabbris}},
  \bibinfo {author} {\bibfnamefont {A.}~\bibnamefont {Said}}, \bibinfo {author}
  {\bibfnamefont {Y.}~\bibnamefont {Pai}}, \bibinfo {author} {\bibfnamefont
  {Q.}~\bibnamefont {Yin}}, \bibinfo {author} {\bibfnamefont {C.}~\bibnamefont
  {Gong}}, \bibinfo {author} {\bibfnamefont {Z.}~\bibnamefont {Tu}}, \bibinfo
  {author} {\bibfnamefont {H.}~\bibnamefont {Lei}}, \bibinfo {author}
  {\bibfnamefont {J.}~\bibnamefont {Sun}}, \bibinfo {author} {\bibfnamefont
  {J.-G.}\ \bibnamefont {Cheng}},  \emph {et~al.},\ }\href@noop {} {\bibfield
  {journal} {\bibinfo  {journal} {arXiv preprint arXiv:2202.13530}\ } (\bibinfo
  {year} {2022})}\BibitemShut {NoStop}%
\bibitem [{\citenamefont {Park}\ \emph {et~al.}(2021)\citenamefont {Park},
  \citenamefont {Ye},\ and\ \citenamefont {Balents}}]{BalentsPRB2021}%
  \BibitemOpen
  \bibfield  {author} {\bibinfo {author} {\bibfnamefont {T.}~\bibnamefont
  {Park}}, \bibinfo {author} {\bibfnamefont {M.}~\bibnamefont {Ye}}, \ and\
  \bibinfo {author} {\bibfnamefont {L.}~\bibnamefont {Balents}},\ }\href
  {\doibase 10.1103/PhysRevB.104.035142} {\bibfield  {journal} {\bibinfo
  {journal} {Phys. Rev. B}\ }\textbf {\bibinfo {volume} {104}},\ \bibinfo
  {pages} {035142} (\bibinfo {year} {2021})}\BibitemShut {NoStop}%
\bibitem [{\citenamefont {Nandkishore}\ \emph {et~al.}(2012)\citenamefont
  {Nandkishore}, \citenamefont {Levitov},\ and\ \citenamefont
  {Chubukov}}]{nandkishore2012chiral}%
  \BibitemOpen
  \bibfield  {author} {\bibinfo {author} {\bibfnamefont {R.}~\bibnamefont
  {Nandkishore}}, \bibinfo {author} {\bibfnamefont {L.~S.}\ \bibnamefont
  {Levitov}}, \ and\ \bibinfo {author} {\bibfnamefont {A.~V.}\ \bibnamefont
  {Chubukov}},\ }\href {https://www.nature.com/articles/nphys2208} {\bibfield
  {journal} {\bibinfo  {journal} {Nature Physics}\ }\textbf {\bibinfo {volume}
  {8}},\ \bibinfo {pages} {158} (\bibinfo {year} {2012})}\BibitemShut {NoStop}%
\bibitem [{\citenamefont {Wang}\ \emph
  {et~al.}(2013{\natexlab{b}})\citenamefont {Wang}, \citenamefont {Li},
  \citenamefont {Xiang},\ and\ \citenamefont {Wang}}]{WanginstPRB2013}%
  \BibitemOpen
  \bibfield  {author} {\bibinfo {author} {\bibfnamefont {W.-S.}\ \bibnamefont
  {Wang}}, \bibinfo {author} {\bibfnamefont {Z.-Z.}\ \bibnamefont {Li}},
  \bibinfo {author} {\bibfnamefont {Y.-Y.}\ \bibnamefont {Xiang}}, \ and\
  \bibinfo {author} {\bibfnamefont {Q.-H.}\ \bibnamefont {Wang}},\ }\href
  {\doibase 10.1103/PhysRevB.87.115135} {\bibfield  {journal} {\bibinfo
  {journal} {Phys. Rev. B}\ }\textbf {\bibinfo {volume} {87}},\ \bibinfo
  {pages} {115135} (\bibinfo {year} {2013}{\natexlab{b}})}\BibitemShut
  {NoStop}%
\bibitem [{\citenamefont {Van~Hove}(1953)}]{VanHove1953}%
  \BibitemOpen
  \bibfield  {author} {\bibinfo {author} {\bibfnamefont {L.}~\bibnamefont
  {Van~Hove}},\ }\href {\doibase 10.1103/PhysRev.89.1189} {\bibfield  {journal}
  {\bibinfo  {journal} {Phys. Rev.}\ }\textbf {\bibinfo {volume} {89}},\
  \bibinfo {pages} {1189} (\bibinfo {year} {1953})}\BibitemShut {NoStop}%
\bibitem [{\citenamefont {Mermin}\ and\ \citenamefont
  {Wagner}(1966)}]{MerminWagnerprl1966}%
  \BibitemOpen
  \bibfield  {author} {\bibinfo {author} {\bibfnamefont {N.~D.}\ \bibnamefont
  {Mermin}}\ and\ \bibinfo {author} {\bibfnamefont {H.}~\bibnamefont
  {Wagner}},\ }\href {\doibase 10.1103/PhysRevLett.17.1133} {\bibfield
  {journal} {\bibinfo  {journal} {Phys. Rev. Lett.}\ }\textbf {\bibinfo
  {volume} {17}},\ \bibinfo {pages} {1133} (\bibinfo {year}
  {1966})}\BibitemShut {NoStop}%
\bibitem [{\citenamefont {Fernandes}\ \emph {et~al.}(2012)\citenamefont
  {Fernandes}, \citenamefont {Chubukov}, \citenamefont {Knolle}, \citenamefont
  {Eremin},\ and\ \citenamefont {Schmalian}}]{Fernandesnematicprb2012}%
  \BibitemOpen
  \bibfield  {author} {\bibinfo {author} {\bibfnamefont {R.~M.}\ \bibnamefont
  {Fernandes}}, \bibinfo {author} {\bibfnamefont {A.~V.}\ \bibnamefont
  {Chubukov}}, \bibinfo {author} {\bibfnamefont {J.}~\bibnamefont {Knolle}},
  \bibinfo {author} {\bibfnamefont {I.}~\bibnamefont {Eremin}}, \ and\ \bibinfo
  {author} {\bibfnamefont {J.}~\bibnamefont {Schmalian}},\ }\href {\doibase
  10.1103/PhysRevB.85.024534} {\bibfield  {journal} {\bibinfo  {journal} {Phys.
  Rev. B}\ }\textbf {\bibinfo {volume} {85}},\ \bibinfo {pages} {024534}
  (\bibinfo {year} {2012})}\BibitemShut {NoStop}%
\bibitem [{\citenamefont {Varma}(2006)}]{VarmaPRB2006}%
  \BibitemOpen
  \bibfield  {author} {\bibinfo {author} {\bibfnamefont {C.~M.}\ \bibnamefont
  {Varma}},\ }\href {\doibase 10.1103/PhysRevB.73.155113} {\bibfield  {journal}
  {\bibinfo  {journal} {Phys. Rev. B}\ }\textbf {\bibinfo {volume} {73}},\
  \bibinfo {pages} {155113} (\bibinfo {year} {2006})}\BibitemShut {NoStop}%
\bibitem [{\citenamefont {Lee}(2014)}]{PALeePRX2014}%
  \BibitemOpen
  \bibfield  {author} {\bibinfo {author} {\bibfnamefont {P.~A.}\ \bibnamefont
  {Lee}},\ }\href {\doibase 10.1103/PhysRevX.4.031017} {\bibfield  {journal}
  {\bibinfo  {journal} {Phys. Rev. X}\ }\textbf {\bibinfo {volume} {4}},\
  \bibinfo {pages} {031017} (\bibinfo {year} {2014})}\BibitemShut {NoStop}%
\bibitem [{\citenamefont {P\'{e}pin}\ \emph {et~al.}(2020)\citenamefont
  {P\'{e}pin}, \citenamefont {Chakraborty}, \citenamefont {Grandadam},\ and\
  \citenamefont {Sarkar}}]{pepin2020fluctuations}%
  \BibitemOpen
  \bibfield  {author} {\bibinfo {author} {\bibfnamefont {C.}~\bibnamefont
  {P\'{e}pin}}, \bibinfo {author} {\bibfnamefont {D.}~\bibnamefont
  {Chakraborty}}, \bibinfo {author} {\bibfnamefont {M.}~\bibnamefont
  {Grandadam}}, \ and\ \bibinfo {author} {\bibfnamefont {S.}~\bibnamefont
  {Sarkar}},\ }\href {\doibase 10.1146/annurev-conmatphys-031218-013125}
  {\bibfield  {journal} {\bibinfo  {journal} {Annual Review of Condensed Matter
  Physics}\ }\textbf {\bibinfo {volume} {11}},\ \bibinfo {pages} {301}
  (\bibinfo {year} {2020})}\BibitemShut {NoStop}%
\bibitem [{\citenamefont {Sarkar}\ \emph {et~al.}(2021)\citenamefont {Sarkar},
  \citenamefont {Grandadam},\ and\ \citenamefont {P\'epin}}]{Sarkarphonon2021}%
  \BibitemOpen
  \bibfield  {author} {\bibinfo {author} {\bibfnamefont {S.}~\bibnamefont
  {Sarkar}}, \bibinfo {author} {\bibfnamefont {M.}~\bibnamefont {Grandadam}}, \
  and\ \bibinfo {author} {\bibfnamefont {C.}~\bibnamefont {P\'epin}},\ }\href
  {\doibase 10.1103/PhysRevResearch.3.013162} {\bibfield  {journal} {\bibinfo
  {journal} {Phys. Rev. Res.}\ }\textbf {\bibinfo {volume} {3}},\ \bibinfo
  {pages} {013162} (\bibinfo {year} {2021})}\BibitemShut {NoStop}%
\bibitem [{\citenamefont {Wang}\ and\ \citenamefont
  {Chubukov}(2014)}]{Wangnematicprb2014}%
  \BibitemOpen
  \bibfield  {author} {\bibinfo {author} {\bibfnamefont {Y.}~\bibnamefont
  {Wang}}\ and\ \bibinfo {author} {\bibfnamefont {A.}~\bibnamefont
  {Chubukov}},\ }\href {\doibase 10.1103/PhysRevB.90.035149} {\bibfield
  {journal} {\bibinfo  {journal} {Phys. Rev. B}\ }\textbf {\bibinfo {volume}
  {90}},\ \bibinfo {pages} {035149} (\bibinfo {year} {2014})}\BibitemShut
  {NoStop}%
\bibitem [{\citenamefont {Tsvelik}\ and\ \citenamefont
  {Chubukov}(2014)}]{tsvelik2014composite}%
  \BibitemOpen
  \bibfield  {author} {\bibinfo {author} {\bibfnamefont {A.}~\bibnamefont
  {Tsvelik}}\ and\ \bibinfo {author} {\bibfnamefont {A.}~\bibnamefont
  {Chubukov}},\ }\href
  {https://journals.aps.org/prb/abstract/10.1103/PhysRevB.89.184515} {\bibfield
   {journal} {\bibinfo  {journal} {Physical Review B}\ }\textbf {\bibinfo
  {volume} {89}},\ \bibinfo {pages} {184515} (\bibinfo {year}
  {2014})}\BibitemShut {NoStop}%
\bibitem [{\citenamefont {Sarkar}\ \emph {et~al.}(2019)\citenamefont {Sarkar},
  \citenamefont {Chakraborty},\ and\ \citenamefont
  {P\'epin}}]{sarkarloopcurrentprb2019}%
  \BibitemOpen
  \bibfield  {author} {\bibinfo {author} {\bibfnamefont {S.}~\bibnamefont
  {Sarkar}}, \bibinfo {author} {\bibfnamefont {D.}~\bibnamefont {Chakraborty}},
  \ and\ \bibinfo {author} {\bibfnamefont {C.}~\bibnamefont {P\'epin}},\ }\href
  {\doibase 10.1103/PhysRevB.100.214519} {\bibfield  {journal} {\bibinfo
  {journal} {Phys. Rev. B}\ }\textbf {\bibinfo {volume} {100}},\ \bibinfo
  {pages} {214519} (\bibinfo {year} {2019})}\BibitemShut {NoStop}%
\bibitem [{\citenamefont {Kim}\ \emph {et~al.}(2023)\citenamefont {Kim},
  \citenamefont {Oh}, \citenamefont {Moon},\ and\ \citenamefont
  {Kim}}]{kim2023monolayer}%
  \BibitemOpen
  \bibfield  {author} {\bibinfo {author} {\bibfnamefont {S.-W.}\ \bibnamefont
  {Kim}}, \bibinfo {author} {\bibfnamefont {H.}~\bibnamefont {Oh}}, \bibinfo
  {author} {\bibfnamefont {E.-G.}\ \bibnamefont {Moon}}, \ and\ \bibinfo
  {author} {\bibfnamefont {Y.}~\bibnamefont {Kim}},\ }\href
  {https://www.nature.com/articles/s41467-023-36341-2} {\bibfield  {journal}
  {\bibinfo  {journal} {Nature Communications}\ }\textbf {\bibinfo {volume}
  {14}},\ \bibinfo {pages} {591} (\bibinfo {year} {2023})}\BibitemShut
  {NoStop}%
\bibitem [{\citenamefont {Song}\ \emph
  {et~al.}(2021{\natexlab{a}})\citenamefont {Song}, \citenamefont {Ying},
  \citenamefont {Chen}, \citenamefont {Han}, \citenamefont {Wu}, \citenamefont
  {Schnyder}, \citenamefont {Huang}, \citenamefont {Guo},\ and\ \citenamefont
  {Chen}}]{Songprl2021}%
  \BibitemOpen
  \bibfield  {author} {\bibinfo {author} {\bibfnamefont {Y.}~\bibnamefont
  {Song}}, \bibinfo {author} {\bibfnamefont {T.}~\bibnamefont {Ying}}, \bibinfo
  {author} {\bibfnamefont {X.}~\bibnamefont {Chen}}, \bibinfo {author}
  {\bibfnamefont {X.}~\bibnamefont {Han}}, \bibinfo {author} {\bibfnamefont
  {X.}~\bibnamefont {Wu}}, \bibinfo {author} {\bibfnamefont {A.~P.}\
  \bibnamefont {Schnyder}}, \bibinfo {author} {\bibfnamefont {Y.}~\bibnamefont
  {Huang}}, \bibinfo {author} {\bibfnamefont {J.-g.}\ \bibnamefont {Guo}}, \
  and\ \bibinfo {author} {\bibfnamefont {X.}~\bibnamefont {Chen}},\ }\href
  {\doibase 10.1103/PhysRevLett.127.237001} {\bibfield  {journal} {\bibinfo
  {journal} {Phys. Rev. Lett.}\ }\textbf {\bibinfo {volume} {127}},\ \bibinfo
  {pages} {237001} (\bibinfo {year} {2021}{\natexlab{a}})}\BibitemShut
  {NoStop}%
\bibitem [{\citenamefont {Song}\ \emph
  {et~al.}(2021{\natexlab{b}})\citenamefont {Song}, \citenamefont {Kong},
  \citenamefont {Xia}, \citenamefont {Yin}, \citenamefont {Tu}, \citenamefont
  {Zhao}, \citenamefont {Dai}, \citenamefont {Meng}, \citenamefont {Tao},
  \citenamefont {Tu} \emph {et~al.}}]{song2021competing}%
  \BibitemOpen
  \bibfield  {author} {\bibinfo {author} {\bibfnamefont {B.}~\bibnamefont
  {Song}}, \bibinfo {author} {\bibfnamefont {X.}~\bibnamefont {Kong}}, \bibinfo
  {author} {\bibfnamefont {W.}~\bibnamefont {Xia}}, \bibinfo {author}
  {\bibfnamefont {Q.}~\bibnamefont {Yin}}, \bibinfo {author} {\bibfnamefont
  {C.}~\bibnamefont {Tu}}, \bibinfo {author} {\bibfnamefont {C.}~\bibnamefont
  {Zhao}}, \bibinfo {author} {\bibfnamefont {D.}~\bibnamefont {Dai}}, \bibinfo
  {author} {\bibfnamefont {K.}~\bibnamefont {Meng}}, \bibinfo {author}
  {\bibfnamefont {Z.}~\bibnamefont {Tao}}, \bibinfo {author} {\bibfnamefont
  {Z.}~\bibnamefont {Tu}},  \emph {et~al.},\ }\href@noop {} {\bibfield
  {journal} {\bibinfo  {journal} {arXiv preprint arXiv:2105.09248}\ } (\bibinfo
  {year} {2021}{\natexlab{b}})}\BibitemShut {NoStop}%
\bibitem [{\citenamefont {Wang}\ \emph {et~al.}(2021)\citenamefont {Wang},
  \citenamefont {Yu}, \citenamefont {Zhang}, \citenamefont {Liu}, \citenamefont
  {Li}, \citenamefont {Peng}, \citenamefont {Di}, \citenamefont {Jiang},\ and\
  \citenamefont {Mu}}]{wang2021enhancement}%
  \BibitemOpen
  \bibfield  {author} {\bibinfo {author} {\bibfnamefont {T.}~\bibnamefont
  {Wang}}, \bibinfo {author} {\bibfnamefont {A.}~\bibnamefont {Yu}}, \bibinfo
  {author} {\bibfnamefont {H.}~\bibnamefont {Zhang}}, \bibinfo {author}
  {\bibfnamefont {Y.}~\bibnamefont {Liu}}, \bibinfo {author} {\bibfnamefont
  {W.}~\bibnamefont {Li}}, \bibinfo {author} {\bibfnamefont {W.}~\bibnamefont
  {Peng}}, \bibinfo {author} {\bibfnamefont {Z.}~\bibnamefont {Di}}, \bibinfo
  {author} {\bibfnamefont {D.}~\bibnamefont {Jiang}}, \ and\ \bibinfo {author}
  {\bibfnamefont {G.}~\bibnamefont {Mu}},\ }\href@noop {} {\bibfield  {journal}
  {\bibinfo  {journal} {arXiv preprint arXiv:2105.07732}\ } (\bibinfo {year}
  {2021})}\BibitemShut {NoStop}%
\bibitem [{\citenamefont {Ratcliff}\ \emph {et~al.}(2021)\citenamefont
  {Ratcliff}, \citenamefont {Hallett}, \citenamefont {Ortiz}, \citenamefont
  {Wilson},\ and\ \citenamefont {Harter}}]{BOrtizPRM2021}%
  \BibitemOpen
  \bibfield  {author} {\bibinfo {author} {\bibfnamefont {N.}~\bibnamefont
  {Ratcliff}}, \bibinfo {author} {\bibfnamefont {L.}~\bibnamefont {Hallett}},
  \bibinfo {author} {\bibfnamefont {B.~R.}\ \bibnamefont {Ortiz}}, \bibinfo
  {author} {\bibfnamefont {S.~D.}\ \bibnamefont {Wilson}}, \ and\ \bibinfo
  {author} {\bibfnamefont {J.~W.}\ \bibnamefont {Harter}},\ }\href {\doibase
  10.1103/PhysRevMaterials.5.L111801} {\bibfield  {journal} {\bibinfo
  {journal} {Phys. Rev. Mater.}\ }\textbf {\bibinfo {volume} {5}},\ \bibinfo
  {pages} {L111801} (\bibinfo {year} {2021})}\BibitemShut {NoStop}%
\bibitem [{\citenamefont {Zhao}\ \emph
  {et~al.}(2021{\natexlab{b}})\citenamefont {Zhao}, \citenamefont {Wu},
  \citenamefont {Wang},\ and\ \citenamefont {Yang}}]{ZhaoDFT2021}%
  \BibitemOpen
  \bibfield  {author} {\bibinfo {author} {\bibfnamefont {J.}~\bibnamefont
  {Zhao}}, \bibinfo {author} {\bibfnamefont {W.}~\bibnamefont {Wu}}, \bibinfo
  {author} {\bibfnamefont {Y.}~\bibnamefont {Wang}}, \ and\ \bibinfo {author}
  {\bibfnamefont {S.~A.}\ \bibnamefont {Yang}},\ }\href {\doibase
  10.1103/PhysRevB.103.L241117} {\bibfield  {journal} {\bibinfo  {journal}
  {Phys. Rev. B}\ }\textbf {\bibinfo {volume} {103}},\ \bibinfo {pages}
  {L241117} (\bibinfo {year} {2021}{\natexlab{b}})}\BibitemShut {NoStop}%
\bibitem [{\citenamefont {Hu}\ \emph {et~al.}(2022{\natexlab{b}})\citenamefont
  {Hu}, \citenamefont {Wu}, \citenamefont {Ortiz}, \citenamefont {Ju},
  \citenamefont {Han}, \citenamefont {Ma}, \citenamefont {Plumb}, \citenamefont
  {Radovic}, \citenamefont {Thomale}, \citenamefont {Wilson} \emph
  {et~al.}}]{hu2022rich}%
  \BibitemOpen
  \bibfield  {author} {\bibinfo {author} {\bibfnamefont {Y.}~\bibnamefont
  {Hu}}, \bibinfo {author} {\bibfnamefont {X.}~\bibnamefont {Wu}}, \bibinfo
  {author} {\bibfnamefont {B.~R.}\ \bibnamefont {Ortiz}}, \bibinfo {author}
  {\bibfnamefont {S.}~\bibnamefont {Ju}}, \bibinfo {author} {\bibfnamefont
  {X.}~\bibnamefont {Han}}, \bibinfo {author} {\bibfnamefont {J.}~\bibnamefont
  {Ma}}, \bibinfo {author} {\bibfnamefont {N.~C.}\ \bibnamefont {Plumb}},
  \bibinfo {author} {\bibfnamefont {M.}~\bibnamefont {Radovic}}, \bibinfo
  {author} {\bibfnamefont {R.}~\bibnamefont {Thomale}}, \bibinfo {author}
  {\bibfnamefont {S.~D.}\ \bibnamefont {Wilson}},  \emph {et~al.},\ }\href
  {https://www.nature.com/articles/s41467-022-29828-x} {\bibfield  {journal}
  {\bibinfo  {journal} {Nature Communications}\ }\textbf {\bibinfo {volume}
  {13}},\ \bibinfo {pages} {2220} (\bibinfo {year}
  {2022}{\natexlab{b}})}\BibitemShut {NoStop}%
\bibitem [{\citenamefont {Lin}\ and\ \citenamefont
  {Nandkishore}(2022{\natexlab{b}})}]{NandkishorePRB2022}%
  \BibitemOpen
  \bibfield  {author} {\bibinfo {author} {\bibfnamefont {Y.-P.}\ \bibnamefont
  {Lin}}\ and\ \bibinfo {author} {\bibfnamefont {R.~M.}\ \bibnamefont
  {Nandkishore}},\ }\href {\doibase 10.1103/PhysRevB.106.L060507} {\bibfield
  {journal} {\bibinfo  {journal} {Phys. Rev. B}\ }\textbf {\bibinfo {volume}
  {106}},\ \bibinfo {pages} {L060507} (\bibinfo {year}
  {2022}{\natexlab{b}})}\BibitemShut {NoStop}%
\bibitem [{\citenamefont {Jos\'e}\ \emph {et~al.}(1977)\citenamefont {Jos\'e},
  \citenamefont {Kadanoff}, \citenamefont {Kirkpatrick},\ and\ \citenamefont
  {Nelson}}]{JKKN1977}%
  \BibitemOpen
  \bibfield  {author} {\bibinfo {author} {\bibfnamefont {J.~V.}\ \bibnamefont
  {Jos\'e}}, \bibinfo {author} {\bibfnamefont {L.~P.}\ \bibnamefont
  {Kadanoff}}, \bibinfo {author} {\bibfnamefont {S.}~\bibnamefont
  {Kirkpatrick}}, \ and\ \bibinfo {author} {\bibfnamefont {D.~R.}\ \bibnamefont
  {Nelson}},\ }\href {\doibase 10.1103/PhysRevB.16.1217} {\bibfield  {journal}
  {\bibinfo  {journal} {Phys. Rev. B}\ }\textbf {\bibinfo {volume} {16}},\
  \bibinfo {pages} {1217} (\bibinfo {year} {1977})}\BibitemShut {NoStop}%
\bibitem [{\citenamefont {Aleiner}\ \emph {et~al.}(2007)\citenamefont
  {Aleiner}, \citenamefont {Kharzeev},\ and\ \citenamefont
  {Tsvelik}}]{AleinerPRB2007}%
  \BibitemOpen
  \bibfield  {author} {\bibinfo {author} {\bibfnamefont {I.~L.}\ \bibnamefont
  {Aleiner}}, \bibinfo {author} {\bibfnamefont {D.~E.}\ \bibnamefont
  {Kharzeev}}, \ and\ \bibinfo {author} {\bibfnamefont {A.~M.}\ \bibnamefont
  {Tsvelik}},\ }\href {\doibase 10.1103/PhysRevB.76.195415} {\bibfield
  {journal} {\bibinfo  {journal} {Phys. Rev. B}\ }\textbf {\bibinfo {volume}
  {76}},\ \bibinfo {pages} {195415} (\bibinfo {year} {2007})}\BibitemShut
  {NoStop}%
\bibitem [{\citenamefont {Anber}\ \emph {et~al.}(2012)\citenamefont {Anber},
  \citenamefont {Poppitz},\ and\ \citenamefont {{\"U}nsal}}]{anber20122d}%
  \BibitemOpen
  \bibfield  {author} {\bibinfo {author} {\bibfnamefont {M.~M.}\ \bibnamefont
  {Anber}}, \bibinfo {author} {\bibfnamefont {E.}~\bibnamefont {Poppitz}}, \
  and\ \bibinfo {author} {\bibfnamefont {M.}~\bibnamefont {{\"U}nsal}},\ }\href
  {https://link.springer.com/article/10.1007/JHEP04(2012)040} {\bibfield
  {journal} {\bibinfo  {journal} {Journal of High Energy Physics}\ }\textbf
  {\bibinfo {volume} {2012}},\ \bibinfo {pages} {1} (\bibinfo {year}
  {2012})}\BibitemShut {NoStop}%
\bibitem [{\citenamefont {Anber}\ \emph {et~al.}(2013)\citenamefont {Anber},
  \citenamefont {Collier},\ and\ \citenamefont {Poppitz}}]{anber20133}%
  \BibitemOpen
  \bibfield  {author} {\bibinfo {author} {\bibfnamefont {M.~M.}\ \bibnamefont
  {Anber}}, \bibinfo {author} {\bibfnamefont {S.}~\bibnamefont {Collier}}, \
  and\ \bibinfo {author} {\bibfnamefont {E.}~\bibnamefont {Poppitz}},\ }\href
  {https://link.springer.com/article/10.1007/JHEP01(2013)126} {\bibfield
  {journal} {\bibinfo  {journal} {Journal of High Energy Physics}\ }\textbf
  {\bibinfo {volume} {2013}},\ \bibinfo {pages} {1} (\bibinfo {year}
  {2013})}\BibitemShut {NoStop}%
\bibitem [{\citenamefont {Lecheminant}(2007)}]{Philippe}%
  \BibitemOpen
  \bibfield  {author} {\bibinfo {author} {\bibfnamefont {P.}~\bibnamefont
  {Lecheminant}},\ }\href {\doibase
  https://doi.org/10.1016/j.physletb.2006.12.079} {\bibfield  {journal}
  {\bibinfo  {journal} {Physics Letters B}\ }\textbf {\bibinfo {volume}
  {648}},\ \bibinfo {pages} {323} (\bibinfo {year} {2007})}\BibitemShut
  {NoStop}%
\bibitem [{\citenamefont {Wu}(1982)}]{wu1982potts}%
  \BibitemOpen
  \bibfield  {author} {\bibinfo {author} {\bibfnamefont {F.-Y.}\ \bibnamefont
  {Wu}},\ }\href {https://link.aps.org/doi/10.1103/RevModPhys.54.235}
  {\bibfield  {journal} {\bibinfo  {journal} {Reviews of modern physics}\
  }\textbf {\bibinfo {volume} {54}},\ \bibinfo {pages} {235} (\bibinfo {year}
  {1982})}\BibitemShut {NoStop}%
\bibitem [{\citenamefont {Francesco}\ \emph {et~al.}(2012)\citenamefont
  {Francesco}, \citenamefont {Mathieu},\ and\ \citenamefont
  {S{\'e}n{\'e}chal}}]{francesco2012conformal}%
  \BibitemOpen
  \bibfield  {author} {\bibinfo {author} {\bibfnamefont {P.}~\bibnamefont
  {Francesco}}, \bibinfo {author} {\bibfnamefont {P.}~\bibnamefont {Mathieu}},
  \ and\ \bibinfo {author} {\bibfnamefont {D.}~\bibnamefont
  {S{\'e}n{\'e}chal}},\ }\href@noop {} {\emph {\bibinfo {title} {Conformal
  field theory}}}\ (\bibinfo  {publisher} {Springer Science \& Business
  Media},\ \bibinfo {year} {2012})\BibitemShut {NoStop}%
\bibitem [{\citenamefont {Tsvelik}\ and\ \citenamefont
  {Kuklov}(2012)}]{tsvelik2012parafermion}%
  \BibitemOpen
  \bibfield  {author} {\bibinfo {author} {\bibfnamefont {A.~M.}\ \bibnamefont
  {Tsvelik}}\ and\ \bibinfo {author} {\bibfnamefont {A.~B.}\ \bibnamefont
  {Kuklov}},\ }\href {\doibase 10.1088/1367-2630/14/11/115033} {\bibfield
  {journal} {\bibinfo  {journal} {New Journal of Physics}\ }\textbf {\bibinfo
  {volume} {14}},\ \bibinfo {pages} {115033} (\bibinfo {year}
  {2012})}\BibitemShut {NoStop}%
\bibitem [{\citenamefont {He}\ \emph {et~al.}(2014)\citenamefont {He},
  \citenamefont {Yin}, \citenamefont {Zech}, \citenamefont {Soumyanarayanan},
  \citenamefont {Yee}, \citenamefont {Williams}, \citenamefont {Boyer},
  \citenamefont {Chatterjee}, \citenamefont {Wise}, \citenamefont {Zeljkovic},
  \citenamefont {Kondo}, \citenamefont {Takeuchi}, \citenamefont {Ikuta},
  \citenamefont {Mistark}, \citenamefont {Markiewicz}, \citenamefont {Bansil},
  \citenamefont {Sachdev}, \citenamefont {Hudson},\ and\ \citenamefont
  {Hoffman}}]{Sachdev2014fermi}%
  \BibitemOpen
  \bibfield  {author} {\bibinfo {author} {\bibfnamefont {Y.}~\bibnamefont
  {He}}, \bibinfo {author} {\bibfnamefont {Y.}~\bibnamefont {Yin}}, \bibinfo
  {author} {\bibfnamefont {M.}~\bibnamefont {Zech}}, \bibinfo {author}
  {\bibfnamefont {A.}~\bibnamefont {Soumyanarayanan}}, \bibinfo {author}
  {\bibfnamefont {M.~M.}\ \bibnamefont {Yee}}, \bibinfo {author} {\bibfnamefont
  {T.}~\bibnamefont {Williams}}, \bibinfo {author} {\bibfnamefont {M.~C.}\
  \bibnamefont {Boyer}}, \bibinfo {author} {\bibfnamefont {K.}~\bibnamefont
  {Chatterjee}}, \bibinfo {author} {\bibfnamefont {W.~D.}\ \bibnamefont
  {Wise}}, \bibinfo {author} {\bibfnamefont {I.}~\bibnamefont {Zeljkovic}},
  \bibinfo {author} {\bibfnamefont {T.}~\bibnamefont {Kondo}}, \bibinfo
  {author} {\bibfnamefont {T.}~\bibnamefont {Takeuchi}}, \bibinfo {author}
  {\bibfnamefont {H.}~\bibnamefont {Ikuta}}, \bibinfo {author} {\bibfnamefont
  {P.}~\bibnamefont {Mistark}}, \bibinfo {author} {\bibfnamefont {R.~S.}\
  \bibnamefont {Markiewicz}}, \bibinfo {author} {\bibfnamefont
  {A.}~\bibnamefont {Bansil}}, \bibinfo {author} {\bibfnamefont
  {S.}~\bibnamefont {Sachdev}}, \bibinfo {author} {\bibfnamefont {E.~W.}\
  \bibnamefont {Hudson}}, \ and\ \bibinfo {author} {\bibfnamefont {J.~E.}\
  \bibnamefont {Hoffman}},\ }\href {\doibase 10.1126/science.1248221}
  {\bibfield  {journal} {\bibinfo  {journal} {Science}\ }\textbf {\bibinfo
  {volume} {344}},\ \bibinfo {pages} {608} (\bibinfo {year}
  {2014})}\BibitemShut {NoStop}%
\bibitem [{\citenamefont {Agterberg}(2008)}]{Agterberg2008}%
  \BibitemOpen
  \bibfield  {author} {\bibinfo {author} {\bibfnamefont {D.~F. . H.~T.}\
  \bibnamefont {Agterberg}},\ }\href {https://www.nature.com/articles/nphys999}
  {\bibfield  {journal} {\bibinfo  {journal} {Nature Physics}\ }\textbf
  {\bibinfo {volume} {4}},\ \bibinfo {pages} {639} (\bibinfo {year}
  {2008})}\BibitemShut {NoStop}%
\bibitem [{\citenamefont {E.Berg}\ and\ \citenamefont
  {Kivelson}(2009)}]{Berg2009}%
  \BibitemOpen
  \bibfield  {author} {\bibinfo {author} {\bibfnamefont {E.~F.}\ \bibnamefont
  {E.Berg}}\ and\ \bibinfo {author} {\bibfnamefont {S.~A.}\ \bibnamefont
  {Kivelson}},\ }\href {https://www.nature.com/articles/nphys1389} {\bibfield
  {journal} {\bibinfo  {journal} {Nature Physics}\ }\textbf {\bibinfo {volume}
  {5}},\ \bibinfo {pages} {830} (\bibinfo {year} {2009})}\BibitemShut {NoStop}%
\bibitem [{\citenamefont {Radzihovsky}\ and\ \citenamefont
  {Ashvin}(2009)}]{Asvin2009}%
  \BibitemOpen
  \bibfield  {author} {\bibinfo {author} {\bibfnamefont {L.}~\bibnamefont
  {Radzihovsky}}\ and\ \bibinfo {author} {\bibfnamefont {V.}~\bibnamefont
  {Ashvin}},\ }\href {https://link.aps.org/doi/10.1103/PhysRevLett.103.010404}
  {\bibfield  {journal} {\bibinfo  {journal} {Phys. Rev. Lett.}\ }\textbf
  {\bibinfo {volume} {103}},\ \bibinfo {pages} {010404} (\bibinfo {year}
  {2009})}\BibitemShut {NoStop}%
\bibitem [{\citenamefont {J.~Ge}\ and\ \citenamefont {Wang}()}]{6e2023}%
  \BibitemOpen
  \bibfield  {author} {\bibinfo {author} {\bibfnamefont {Y.~X. Q. Y. H. L.
  Z.~W.}\ \bibnamefont {J.~Ge}, \bibfnamefont {P.~Wang}}\ and\ \bibinfo
  {author} {\bibfnamefont {J.}~\bibnamefont {Wang}},\ }\href@noop {} {\bibinfo
  {journal} {arXiv preprint arXiv: 2201.10352}\ }\BibitemShut {NoStop}%
\end{thebibliography}%
\end{document}